\documentclass[useAMS,usenatbib]{mnras}
\usepackage[fleqn]{amsmath}
\usepackage{graphicx}	% Including figure files
\pdfoutput=1 
\newcommand{\Mz}{M_{\rm s0}} 
\newcommand{\Rhz}{R_{\rm h0}} \newcommand{\Mhz}{M_{\rm h0}}
\newcommand{\der}{{\rm d}}

 \newcommand{\tang}{_{\rm t}}
 
 \newcommand{\acc}{^{\rm acc}}

 \newcommand{\h}{_{\rm h}} 
\newcommand{\mini}{^{\rm min}} 
 
\newcommand{\ta}{^{\rm ntr}} \newcommand{\clR}{R_{\rm s}} 
\newcommand{\clM}{M_{\rm s}} 
\newcommand{\clQ}{Q_{\rm s}}  
 
\newcommand{\clc}{c_{\rm s}}  
 \newcommand{\per}{_{\rm per}}
 \newcommand{\tr}{^{\rm tr}} \newcommand{\trb}{_{\rm rel} }
\newcommand{\trh}{^{\rm tr,h}} 
\newcommand{\trs}{^{\rm tr,s}} 
 
 \newcommand{\des}{_{\rm dis}}
 \newcommand{\dc}{_{\rm dDM}} 
 \newcommand{\res}{_{\rm min}} 
 \newcommand{\s}{_{\rm s}}

\newcommand{\maxi}{_{\rm max}} 
\newcommand{\modot}{M$_\odot$\ } \newcommand{\modotc}{M$_\odot$}
\newcommand{\ti}{t_{\rm i}} 
\newcommand{\ii}{_i}

\newcommand{\beq}{\begin{equation}} \newcommand{\eeq}{\end{equation}}
 \newcommand{\beqa}{\begin{eqnarray}}
\newcommand{\eeqa}{\end{eqnarray}} \newcommand{\lav}{\langle}
\newcommand{\rav}{\rangle}

 \newcommand{\clf}{f_{\rm s}}
\newcommand{\fin}{^{\rm stp}}

\begin{document}

\title[II. Stripped Subhaloes]{An Accurate Comprehensive Approach to Substructure:\\ II. Stripped Subhaloes}
%{Unveiling the Origin of the Typical Properties of Halo Substructure}

\author[Salvador-Sol\'e, Manrique \& Botella]{Eduard
  Salvador-Sol\'e$^1$\thanks{E-mail: e.salvador@ub.edu}, Alberto Manrique$^1$ and Ignacio Botella$^{1,2}$
  \\$^1$Institut de Ci\`encies del Cosmos, Universitat de
  Barcelona, Mart{\'\i} i Franqu\`es 1, E-08028 Barcelona, Spain
  \\$^2$Dept. of Astronomy, Graduate School of Science, Kyoto University, Kitashirakawa, Oiwakecho, Sakyo-ku, Kyoto, 606-8502, Japan}

%% Abstract and keywords

\maketitle
\begin{abstract}
In Paper I we used the CUSP formalism to derive from first principles and no single free parameter the accurate abundance and radial distribution of both diffuse DM (dDM) and subhaloes accreted onto haloes and their progenitors at all previous times. Here we use those results as initial conditions for the monitoring of the evolution of subhaloes and dDM within the host haloes. Specifically, neglecting dynamical friction, we accurately calculate the effects of repetitive tidal stripping and heating on subhaloes as they orbit inside the host halo and infer the amount of dDM and subsubhaloes they release into the intra-halo medium. We then calculate the expected abundance and radial distribution of stripped subhaloes and dDM. This derivation clarifies the role of halo concentration in substructure and unravels the origin of some key features found in simulations including the dependence of substructure on halo mass. In addition, it unveils the specific effects of dynamical friction on substructure. The results derived here are for purely accreting haloes. In Paper III we complete the study by addressing the case of low-mass subhaloes, unaffected by dynamical friction, in ordinary haloes having suffered major mergers.
\end{abstract}


%% From the front matter, we move on to the body of the paper.

\section{INTRODUCTION}\label{intro}

Halo substructure is a subject of paramount importance for its multiple implications in many astrophysical issues. It has thus been amply studied by all available means.

High-resolution $N$-bdy simulations (e.g. \citealt{Dea07}; \citealt{Sea08a}, hereafter SWV; \citealt{Aea09,Eea09,BK10,Gi10,Kea11,Gea11,Gea12,Oea12,Lea14,Cea14,Iea20}) and, more recently, hydrodynamical simulations (\citealt{Rich20,Fea20,Fea20b}; see also \citealt{Hell16,Bea16,Bea20} for the inclusion of gas using a semi-analytic treatment) have allowed to characterise their properties, while analytic models have been used to try to understand the origin of those properties (e.g. \citealt{TB01,Fea02,ZB03,S03,L04,OL04,TB04,Pe05,vdB05,Zeabis05,KB07,Gi08,Bea13,PB14,Ji16,Gfea16,vdB16}). 

The modelling of substructure is particularly hard. One must account for the rate at which haloes of different masses are accreted onto the host halo and converted into subhaloes, determine their initial radial and velocity distributions and monitor their fate as they orbit inside the host halo. Subhaloes neither accrete gas (all cooled gas goes to the centre of the halo) nor merge with each other (they have too large relative velocities; \citealt{Aea09}) nor even suffer significant harassment \citep{vdB18}. But they are tidally stripped and heated by the potential well of the host halo in a complicate way that depends not only on their initial location and velocity, but also on their varying mass and concentration. In addition, the host haloes themselves have different assembly histories, which translates into different evolving histories of subhaloes. It is thus not unsurprising that, despite all the efforts gone on this study using both (semi)analytic models plus numerical experiments (e.g. \citealt{vdB16,vdB18,GB19,Jea21}) and simulations (e.g. \citealt{Gea98,Hay03}), the origin of the characteristic properties of substructure remains an open issue.

One interesting result along this line of research was obtained by \citet{Han15}, hereafter HCFJ. These authors showed that the properties of subhaloes found in simulations are encoded in the three following conditions: 1) the scaled number density profiles of subhaloes with original mass $\clM$ that were accreted onto the halo or its progenitors at all previous times overlap in one curve proportional to the scaled density profile of the host halo; 2) the cumulative mass function (MF) of such accreted subhaloes is a power law with logarithmic slope, $\der {\cal N}(> \clM)/\der \ln \clM$, close to $-1$; and 3) the truncated-to-original mass ratio of the final stripped subhaloes only depends on their radial distance to the centre of the halo. Strictly speaking, the two first conditions refer to `unevolved' rather than `accreted' subhaloes, but both kinds of subhaloes coincide for subhaloes with low enough masses unaffected by dynamical friction (see Paper I). Unfortunately, what causes these conditions is unknown. 

With the aim to shed light on this issue, a novel approach was applied in Paper I \citep{I} making use of the so-called {\it ConflUent System of Peak trajectories} (CUSP) formalism \citep{Mea95,Mea96,Mea98}. CUSP is a powerful analytic formalism that, by monitoring the collapse and virialisation of halo seeds, i.e. peaks (or maxima) in the Gaussian random field of density perturbations \citep{Jea14a}, allows one to infer {\it from first principles and with no single free parameter} all macroscopìc halo properties, namely the mass function (MF) \citep{Jea14b} and the mean spherically averaged density \citep{Sea12a}, velocity dispersion and anisotropy profiles as well as the prolateness and ellipticity profiles (\citealt{Sea12b}, hereafter SSMG). Not only are the predictions in full agreement with the results of simulations in both CDM (all previous references) as well WDM \citep{Vea12} cosmologies, but their derivation clarifies the origin of all these properties and their characteristic features. A comprehensive review of CUSP and its achievements is given in \citet{SM19}.

In Paper I we extended the domain of application of CUSP to the basic halo components: diffuse dark matter (dDM) and subhaloes. We derived the MF and radial distribution of accreted subhaloes which allowed us to explained the origin of the two first above mentioned HCFJ conditions. In the present Paper we use those results as initial conditions for the study of the fate of dDM and subhaloes within the host haloes. For simplicity, we concentrate on haloes evolving by pure accretion, which is enough to explain the third HCFJ condition. The more realistic though complicated case of ordinary haloes having suffered major mergers is postponed to Paper III \citep{III}. 

The layout of the Paper is as follows. In Section \ref{peaks}, we remind the results of Paper I. In Section \ref{stripping} we carefully model tidal stripping of subhaloes by the host potential well. The abundance and radial distribution of stripped subhaloes and of diffuse dark matter (dDM) are derived in Sections \ref{stripped} and \ref{dDM}, respectively, under the approximation that all subhaloes have similar concentrations. In Section \ref{masses} we relax that approximation and analyse the dependence of substructure on halo mass. Our results are summarized and discussed in Section \ref{dis}.

Throughout the Paper our predictions are calculated for current Milky Way (MW)-like haloes with virial mass, i.e. the mass out to the radius encompassing an inner mean density equal to the virial overdensity \citep{BN98,H00} times the mean cosmic density, of $M\h=2.2\times 10^{12}$ \modotc, which according to \citet{SM19} correspond to the maximum extend of the virialised part of haloes. These predictions are compared to the results for the same kinds of haloes studied by HCFJ or SWF, who use the masses $M_{200}=1.84\times 10^{12}$ \modot or $M_{50}=2.5\times 10^{12}$ \modotc, respectively, i.e. out to the radius encompassing an inner mean density of 200 or 50 times the critical cosmic density. The cosmology adopted is the {\it WMAP7} cosmology \citep{Km11} as in those latter works. The CDM spectrum we use is according to the prescriptiobn given by \citet{BBKS} with the \citet{S95} shape parameter.

Given that in the present paper we deal with both stripped and accreted subhaloes, all properties referring to the former are denoted with superindex `stp', whereas those referring to the latter, derived in Paper I, are denoted with superindex `acc'. 

\section{ACCRETED SUBHALOES}\label{peaks}

The main results of Paper I are summarised next (see that Paper for details). 

\begin{itemize}

\item There is a one-to-one correspondence between haloes with mass $M\h$ at the time $t\h$ and their seeds, non-nested peaks with density contrast $\delta$ at the scale $S$, in the Gaussian random field of density perturbations at an arbitrary initial time $\ti$ smoothed with a Gaussian window. See Paper I for the functions $\delta(t\h)$ and $S(M\h,t\h)$ corresponding to the cosmology and halo mass definition of interest. 

\item Consequently, the continuous time evolution of accreting haloes is traced by continuous peak trajectories in the $\delta$--$S$ plane at $\ti$. Those continuous trajectories are interrupted in major mergers, where new continuous peak trajectories arise tracing the evolution of the haloes arising from the mergers. 

\item The previous correspondence yields in turn another one-to-one correspondence between subhaloes of mass $\clM$ accreted by the halo (or any of its progenitors) at any time $t\le t\h$ and peaks with $\delta(t)$ at the scale $S(\clM)$ that become nested in the peak with the same density contrast at a larger scale associated with the host halo.

\item When a halo is accreted onto another halo and becomes a subhalo, its associated peak becomes nested in the peak tracing the host halo. On the other hand, nested peaks are preserved like subhaloes when haloes suffer major mergers. Thus, the dynamical evolution of subhaloes of any level can be monitored through the filtering evolution of nested peaks of the same level in the $\delta$--$S$ plane. 

\item For the reasons explained in \citet{SM19}, the properties of haloes including those regarding accreted subhaloes do not depend on their assembly history. In other words, they do not depend on whether haloes undergo monolithic collapse (pure accretion) or lumpy collapse (including major mergers). Consequently, to derive all these properties we have the right to assume pure accretion with no loss of generality.

\item During accretion phases haloes grow inside-out, i.e. shells accreted at a time $t(r)$ when the halo reached the mass $M(r)$,\footnote{$t(r)$ is given by the trajectory $\delta(S)$ tracing the halo growth, with the relations $\delta(t)$ and $S(M\h,t)$ defining the halo-peak correspondence, and the mass profile $M(r)$ of the halo (see Paper I).} are deposited at the radius $r$ without altering the inner structure of the halo. By `deposited at $r$' we mean that the subhalo orbits stabilise with their apocentre at that radius. 

\item The orbits of subhaloes accreted at $t(r)$ are determined by their random (tangential) velocities $v$ at their apocentre at $r$, which arise from the collapse and virialisation of the halo,\footnote{The virial relation we refer to throughout this Paper includes the external pressure term, so by virialised haloes we simply mean relaxed ones.} so they do not depend on the subhalo mass $\clM$.  
\item The inside-out growth of purely accreting haloes allows one to derive their mean spherically averaged density profile as well as the abundance and mean spherically averaged number density profile per infinitesimal mass of accreted subhaloes from the abundance of nested peaks arising from Gaussian statistics.

\item CUSP also accounts for the dDM outside haloes arising from the existence of a minimum halo mass due to the free-streaming mass associated to WIMPs in the real Universe or the halo resolution mass in simulations. When this dDM is accreted onto haloes, it gives rise to a non-null dDM mass fraction, $f\dc\acc(r)$, at each radius $r$ of haloes. 

\end{itemize}

\section{TIDAL STRIPPING AND HEATING}\label{stripping}

To infer the properties of substructure regarding stripped subhaloes, we must first determine the effects of tidal strippping and heating on accreted subhaloes orbiting inside the accreting host halo. For simplicity in the calculation of subhalo orbits, we will assume all objects spherically symmetric. This makes a difference with respect to Paper I. In that Paper the results obtained, in the form of mean spherically averaged profiles, held for real haloes with different ellipsoidal shapes. Here, instead, the results we will obtain strictly hold for spherically symmetric systems only. Nonetheless, they should hopefully be a good approximation for all haloes duly spherically averaged.

\subsection{Truncation}\label{single}

The truncated mass $\clM\tr(v,r,\clM)$ of subhaloes with original mass $\clM$ can be calculated from their density profile (see below) and the truncation radius $\clR\tr(v,r,\clM)$. 

At present there is no consensus about how to estimate the truncation radius (see the detailed discussion in \citealt{vdB18}). The most usual procedure is to adopt the tidal-limited radius of the subhalo in a circular orbit at $r$, i.e. the radius of the subhalo in equilibrium within the tidal field of the host halo at that radius. But even this can be done in several ways, depending on whether or not the centrifugal force (e.g. \citealt{K62,Sp87,Tea98,BT08,Tea17}) or resonant effects \citep{Kea99} are taken into account, none of them being fully accurate \citep{BT08,Rea06,Mea10}. However, we do not need any accurate value of the truncation radius. For our purposes here it is sufficient to consider that all the previous procedures lead to similar results: the truncation radius encompasses an inner mean density of the subhalo at $r$, $\bar\rho_{[\clM,t(r)]}$, of the order of the mean inner density of the host at radius $r$, $\bar\rho(r)$ (e.g. \citealt{Hay03,Dea07,Pea08}),
\beq
\bar\rho_{[\clM,t(r)]}[\clR(r,\clM)]\approx \bar\rho(r)\,.
\label{3}
\eeq
To write equation (\ref{3}) we have used that accreted subhaloes with apocentre at $r$ were accreted there so that the truncation radius is but the subhalo original radius $\clR$ at accretion. Hereafter, a bar on a function of $r$ denotes the corresponding mean value inside that radius and subindex $[M,t]$ on a (sub)halo property means that the object has a mass $M$ at the time $t$. However, in the case of the accreting host halo we drop, for simplicity, the subindex $[M\h,t\h]$ or $[M(r),t(r)]$. 

However, this is not the whole story because subhalo orbits are not circular but elliptical in general. Thus, the truncation radius of subhaloes varies over their orbit, from the one set at the apocentric radius $r$ where stripping is less intense, well-approximated by the relation (\ref{3}), to that set at the pericentric radius $r\per$ where stripping is maximum. In fact, given that subhaloes have large velocities at pericentre, to calculate the truncation radius there it is preferable to use the impulse approximation (\citealt{Sp58}). But, again, we do not need to derive the accurate truncation radius according to the mass loss calculated in the impulsive approximation (e.g. \citet{GO99,vdB18}). For our purposes here it is enough the result found by \citet{Gea94} that $\clR\tr(v,r,\clM)$ also encompasses a mean inner density in the subhalo of the order of that of the host at $r\per(v,r)$,
\beq
\bar\rho_{[\clM,t(r)]}[\clR(r,\clM) \clQ(v,r,\clM)]\approx\,\bar\rho[r_{\rm per}(v,r)]\,,
\label{4}
\eeq
where $\clQ(v,r,\clM)$ stands for their scaled truncated radius, $\clR\tr(v,r,\clM)/\clR(r,\clM)$. We remark that the truncation in the impulsive approximations is not directly due to the local tidal field, but to the heating produced in the subhalo at its rapid passage by pericentre, which causes a more marked stripping to the subhalo. This is why this process is often referred as `tidal heating' or `shock heating'. In the present work we call `truncation' the stripping produced in any of these two extreme ways, via tidal limiting radius or via shock heating, (or any mixture of them) and call `heating' the small energy increase that affects the part of the subhalo that remains bound after being truncated. 

After being stripped and heated at pericentre, all over their way back to the apocentre, subhaloes tend to reach a new equilibrium state limited by the tidal field at each point. As a consequence, subhaloes can be stripped not only in the first half of their orbits, with an ever increasing tidal field, but possibly also in the second half (this depends on how quick is the response of the system to any previous truncation). However, as the stripping is maximum and much stronger via shock heating at pericentre than via tidal limitation at any other point of the orbit, we adopt for simplicity the viewpoint that the whole truncation (and heating) is concentrated at the pericentre and that, at apocenre, subhaloes just re-accomodate their structure according to the halo mean inner density there. We remark, however, that this simplification should have no practical consequence because, as shown below, we will make sure that the total mass loss produced in any orbit is according to the results of simulations.

Equations (\ref{4}) and (\ref{3}) lead to
\beq 
\frac{\bar\rho_{[\clM,t(r)]}[\clR(r,\clM) \clQ(v,r,\clM)]}{\bar\rho_{[\clM,t(r)]}[\clR(r,\clM)]}\approx 
\frac{\bar\rho[r\per(v,r)]}{\bar\rho(r)}.
\label{first}
\eeq
Lastly, using the expression 
\beq
\frac{\clM\tr(v,r,\clM)}{\clM}=\frac{f\left[\clc(r) \clQ(v,r,\clM)\right]}{f[\clc(r)]}
\label{second}
\eeq 
holding for subhaloes with the original NFW density profile \citep{NFW97} with concentration $\clc(r)$ and the similar expression for the mass ratio $M[r\per(r,v)]/M\h$ holding for the host, the relation (\ref{first}) takes the form
\beq
\frac{f[\clc(r) \clQ(v,r,\clM)]}{f[\clc(r) ][\clQ(v,r,\clM)]^3}= 
\frac{f\left[c(r)Q(v,r)\right]}{f[c(r)][Q(v,r)]^3},
\label{mtr2}
\eeq
where $Q(v,r)$ stands for $r\per(v,r)/r$ and $c(r)$ is the concentration of the accreting halo with mass $M(r)$, and $f(c)$ is defined as $\ln(1+c)-{c}/(1+c)$ (see e.g. \citealt{Jea19} for the Einasto profile \citep{E65}). For simplicity, the uncertainty factor of order unity has been taken equal to one. 

Once the truncation radius or, equivalently, the ratio $\clQ(v,r)$ is known, equation (\ref{second}) gives the truncated mass $\clM\tr(v,r,\clM)$. Note that, for subhaloes accreted at $r$ with similar concentrations $\clc(r)$, equation (\ref{mtr2}) implies that $\clQ$ does not depend on $\clM$, i.e. $\clQ(v,r,\clM)\equiv \clQ(v,r)$. Consequently, the mass ratio $\clM\tr/\clM$ given in equation (\ref{second}) and its mean or median value over $v$ are independent of $\clM$. 

\subsection{Heating}

But things are not that simple. When subhaloes settle in a new equilibrium state at apocentre, their density profiles adopt again the NFW form (SWV) with a somewhat larger concentration $\clc\tr$ due to the heating produced at pericentre. Therefore, they will be further stripped and heated at the next orbit and so on so forth. In other words, stripping and heating is a repetitive process. 

To calculate the mass loss produced in the next orbit we must determine the new concentration $\clc\tr$, which in turn depends on the heating produced in the previous orbit. In the impulsive approximation, such a heating mostly affects the outer regions of the subhalo, i.e. those which are precisely lost, while the energy of the innermost regions is rather an adiabatic invariant \citep{Sp87,W94,GO99,vdB18} so that the energy gain of the non-truncated part of the system is quite limited. Taking into account the form of the total energy for subhaloes with the NFW density profile, we then arrive at the expression
\beq
\frac{h[\clc\tr(v,r)]}{h[\clc(r)]}=U(v,r) \left[\frac{\clM\tr(v,r,\clM)}{\clM}\right]^{5/6},
\label{rat2} 
\eeq
where $h(c)$ is defined as $f(c)(1+c)/\{c^{3/2}[3/2-s^2(c)]^{1/2}\}$, being $s^2(c)$ the isotropic 3D velocity variance scaled to $cf(c)GM/R$ of a halo with mass $M$, radius $R$ and concentration $c$, and $U(v,r)$ is the ratio between the total energies after and before the tidal shock of the part of the subhalo {\it that remains bound}. Thus $U(v,r)$ should be a function of order unity of the strength of the shock or, equivalently, of the subhalo truncated-to-original mass ratio. But the adiabatic shielding against heating in the inner part of subhaloes suffering the shock makes it hard to calculate it accurately. \citet{vdB18} provide an approximate expression for the ratio between the total energies of the subhalo after and before the shock, but those energies refer to the whole system, not to the part that remains bound. We thus assume, for simplicity, that in the relevant range of subhalo orbits $U(v,r)$ can be approximated by a power-law,
\beq
U(v,r)=K \left[\frac{\clM\tr(v,r,\clM)}{\clM}\right]^{\beta}.
\label{rat3} 
\eeq
The positive constant of order one $K$ and the negative index with small absolute value $\beta$ will be adjusted below by comparing the predictions of this model to the results of numerical simulations. Note that, unless $K$ is exactly equal to unity, the approximation (\ref{rat3}) cannot be valid for orbits close to circular because $\clM\tr(v,r,\clM)/\clM$ and $U(v,r)$ should then approach unity in parallel. Fortunately, subhaloes with nearly circular orbits are very rare \citep{T97,Zeabis05,W11}, so this slight flaw of the model should have a negligible effect in the results. In any event, disregarding its exact form, the function $U$ cannot depend on $\clM$ because $\clM\tr(v,r,\clM)/\clM$ does not, so $\clc\tr/\clc$ does not depend on $\clM$ either (see eq.~[\ref{rat2}]).

\subsection{Repetitive Stripping} 

To calculate the final truncated mass of subhaloes with original mass $\clM$ accreted at $t(r)$ onto a purely accreting halo with $M\h$ at $t\h$ we must monitor the changes produced in subhaloes at every pericentric passage between $t(r)$ and $t\h$ in an iterative way. The result will depend of course on the concentration $c(r)$ of the host halo at $t(r)$ inside which subhaloes accreted at that time orbit and on the initial properties of those subhaloes. 

Given the inside out growth of the host halo, $c(r)$ is simply $r/r\s$, where $r\s$ is the core radius of the halo at the final time $t\h$. Regarding subhaloes, after turnaround they fall onto the halo and start orbiting and being stripped and heated, so their mass $\clM$ when their apocentre is stabilised at $r$ is somewhat smaller than the mass $\clM\ta$ they would have had they evolved as free (non-accreted) haloes until $t(r)$. Since during virialisation subhalo velocities vary randomly independently of their mass, all subhaloes at $r$ must have the same velocity distribution \citep{Jea15}. Thus we can derive their typical scaled truncation radius $\clQ$ from equation (\ref{mtr2}) from the initial subhalo concentration $\clc\ta$ given by the $M$--$c$ relation for $\clM\ta$ at $t(r)$ and the function $Q(r)$ equal to twice the velocity-average value obtained of subhaloes at $r$ and $t(r)$.\footnote{According to the spherical collapse model \citep{BN98} justified by CUSP \citep{SM19}, subhalo apocentric radii typically shrink a factor two since turnaround, so do also their pericentric radius $r\per$, implying that $r\per/r$ for subhaloes at $r$ was typically a factor 2 larger before being accreted.} Then, plugging the solution $\clQ$ in equation (\ref{second}), we obtain $\clM/\clM\ta$ and, using equations (\ref{rat2})--(\ref{rat3}), we are led to the concentration $\clc$ of accreted subhaloes. 

According to the results of Sec.~\ref{single}, provided all haloes with different masses $\clM\ta$ accreted at $t(r)$ had similar concentrations $\clc\ta$, the mass ratio $\clM/\clM\ta$ would be a function independent of $\clM\ta$, hereafter denoted as $m(r)$,
and so would also $\clQ$ and $\clc$. The derivation followed next makes that approximation in order to catch the main effect of subhalo stripping; the slight dependence on subhalo concentration on mass will be addressed in Section \ref{masses}. Since this derivation uses an iterative procedure, it is convenient to denote the initial properties $\clM(r,\clM\ta)$, $\clQ(r)$ and $\clc(r)$ of accreted subhaloes as $M_0(r,\clM\ta)$ (with $M_0(r,\clM\ta)< M(r)/3$; see Paper I), $Q_0(r)$ and $c_0(r)$, respectively, and increase the subindex in one unit at each new orbit.

\begin{figure*}
\centerline{\includegraphics[scale=0.4,bb= 210 0 872 480]{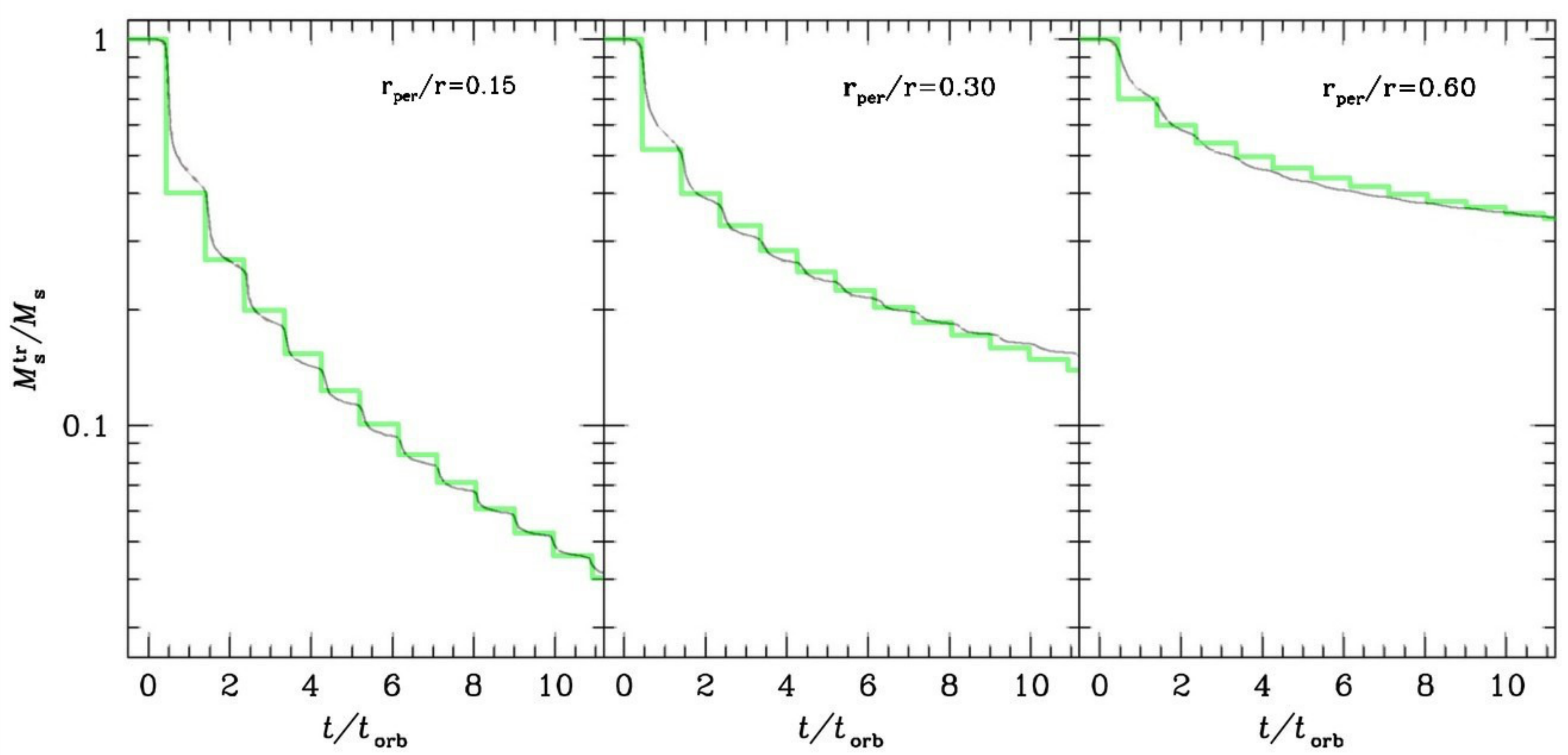}}
\caption{Predicted mass evolution of subhaloes with the NFW density profile and suited concentrations $\clc$ (green lines) in repetitive stripping within a halo with the same density profile and concentration $c(r)=10$ for three different values of $r\per(v,r)/r$, compared to the results of a dedicated numerical experiment (thin black lines) performed by \citet{Hay03}. The smoother shape of the empirical curves is due to the actual progressive stripping and response of subhaloes to the conditions found over the orbits, while the theoretical predictions focus on the maximum stripping taking place at the pericentre and the recovery of equilibrium of the system at apocentre.}
(A colour version of this Figure is available in the online journal.)
\label{f1}
\end{figure*}
 
At first pericentric passage after $t(r)$, subhaloes are truncated at the scaled truncation radius $Q_1(v,r)$ given by equation (\ref{mtr2}) and acquire the mass $M_1(v,r,\clM\ta)$ given by equation (\ref{second}) or, taking into account equation (\ref{mtr2}) also holding for $M_0/\clM\ta$, by the relation
\beq
\frac{M_1(v,r,\clM\ta)}{M_0(r,\clM\ta)}=\frac{M_0(r,\clM\ta)}{\clM\ta}\left[\frac{Q_1(v,r)}{Q_0(r)}\right]^3\,.
\label{third}
\eeq 
As $M_0/\clM\ta$ does not depend on $\clM\ta$ (see the discussion above), $M_1/M_0$ does not either. Then, equation (\ref{rat2}) leads to the new concentration $c_1(v,r)$ when subhaloes reach apocentre. 

Similarly, at second and successive passages, subhaloes with $M\ii$, $Q\ii$ and $c\ii$ are truncated at the scaled truncation radius $Q_{i+1}(v,r)$ given by
\beq
\frac{f[c\ii(v,r) Q_{\rm i+1}(v,r)]}{f[c\ii(v,r) ]Q^3_{\rm i+1}(v,r)}= 
\frac{f\left[c(r)Q(v,r)\right]}{f[c(r)]Q^3(v,r)},
\label{mtr2bis}
\eeq
leading to a mass $M_{i+1}$ satisfying 
\beq 
\frac{M_{i+1}(v,r,\clM\ta)}{M\ii(v,r,\clM\ta)}=\frac{M\ii(v,r,\clM\ta)}{M_{i-1}(v,r,\clM\ta)}\left[\frac{Q_{i+1}(v,r)}{Q\ii(v,r)}\right]^3\
\label{fourth}
\eeq
and to a concentration $c_{i+1}$ according to
\beq
\frac{h[c_{i+1}(v,r)]}{h[c_i(v,r)]}=K\left[\frac{M_{i+1}(v,r,\clM\ta)}{M_i(v,r,\clM\ta)}\right]^{\beta+5/6}.
\label{ratot3} 
\eeq

This iterative process leads to a total truncated mass $\clM\tr$ at $t\h$ of subhaloes with original mass $M_0=\clM$ and scaled truncation radius $Q_0=\clQ$ equal to
\beq
\frac{\clM\tr(v,r,\clM\ta)}{\clM(r,\clM\ta)}=m^\nu (r)\prod_{i=1}^{\nu}\left[\frac{Q_i(v,r)}{\clQ(r)}\right]^3\,,
\label{last}
\eeq
where $\nu>0$ is the total number of constant orbits achieved by subhaloes from $t(r)$ to $t\h$. 

As shown in Figure \ref{f1}, for $K=0.77$ and $\beta=-1/2$, the predictions of the model for a wide range of orbits recover the results of numerical simulations by \citet{Hay03}. As the best value of $K$ is smaller than unity, to avoid an artificial cooling in nearly circular orbits, i.e. when $\clM\tr/\clM$ approaches unity, we take from now on the expression (\ref{rat3}) bounded to unity.

Equation (\ref{last}) tells that the truncated-to-original mass ratio $\clM\tr/\clM$ does not depend on $\clM$ and the same is true for its mean (or median) value over $v$. We thus see that the reason for HCFJ condition 3 in the realistic case of repetitive stripping is the same as in single orbits: the {\it similar} concentration of subhaloes at accretion and the fact that subhaloes at $r$ are truncated at the radius encompassing a mean inner density of the order of that of the host halo at $r$. 

\subsection{Disruption}\label{destruction}

The possibility that subhaloes can be fully disrupted as a consequence of tidal stripping is not clear. \citet{Hay03} pointed out that the total energy of subhaloes endowed with a NFW density profile can become positive after truncation so that they can be disrupted. However, \citet{vdB18} showed that, when a subhalo is being stripped, its structure quickly responds to the departure from equilibrium so that the total energy of the new truncated system is not simply that of the part of the initial system out to the effective truncation radius. Moreover, even if the system could not immediately respond to stripping, the total energy of a severely truncated subhalo with the NFW profile would only become positive provided its velocity distribution were strongly tangential, while in real (sub)haloes it is radially biased. 

According to these arguments, the full disruption of subhaloes would be a very rare event. However, numerical simulations do find a significant disruption above the subhalo resolution mass. Whether it is due to overmerging or any other numerical artifact as claimed by \citet{vdB18} is not clear. In any event, if we are to compare our theoretical predictions to the results of simulations, we must account for it. Therefore, we will consider two scenarios: one with negligible disruption and another one with disruption, less marked than in the isotropic case though. The rest of this section is devoted to this latter scenario.

The pericentric radius $r\per\equiv rQ(v,r)$ of a subhalo is related to its tangential velocity $v$ at the apocentre through
\beq
v^2=2\left\{\frac{\Phi(r)-\Phi[rQ(v,r)]}{1-Q^2(v,r)}\right\},
\label{v-rper}
\eeq
where $\Phi(r)=-GM(r)/ r\,\ln [1+c(r)]/f[c(r)]$ is the potential of the host at $r$. To leading order, equation (\ref{v-rper}) leads to 
\beq
Q(v,r)=\frac{1}{2}\frac{rv^2}{GM(r)}.
\label{v-Q}
\eeq

Since accreted subhaloes become increasingly resistant to disruption due to the increase of their concentration at each passage by the pericentre, they can only be disrupted at first pericentre passage when their concentration is $c_0(r)$. Assuming that the condition for disruption for subhaloes with concentration $\clc$ is $R\tr/R \equiv \clQ \sim  0.5/\clc$,\footnote{The critical value for disruption found by \citealt{Hay03} for isotropic NFW subhaloes is $0.77/\clc$.} equation (\ref{mtr2}), with $\clc=c_0(r)$ and $\clQ$ given by equation (\ref{v-Q}), leads to a tangential velocity for disruption at $r$, $v\des(r)$, satisfying the relation
\beq
\frac{f[c_0(r)]}{f(0.5)}[1+c(r)]^2\left[\frac{r v\des^2(r)}{2GM(r)}\right]^3- c^2(r)\frac{r v\des^2(r)}{2GM(r)}-1=0
\label{vdes}
\eeq

We thus see that $v\des(r)$ is independent of $M_0$ (and $\clM$). From equations (\ref{v-Q}) and (\ref{vdes}) we also have that the minimum pericentric radius for surviving subhaloes is
\beq
r\per\mini= r\, \frac{f[c_0(r)]}{f(0.5)}\left[\frac{0.5}{c_0(r)}\right]^3.
\label{rpermin}
\eeq

The fraction of accreted subhaloes with original mass $\clM$ that are destroyed, $f\des(r,\clM)$, is thus equal to the integral of velocity distribution function, ${\cal N}(v,r,\clM)$, up to $v\des(r)$ divided by the same integral up to $v\maxi(r)=[GM(r)/r)]^{1/2}$, i.e. the maximum possible value of $v$ for subhaloes with apocentre at $r$. For any reasonable (mass-independent) tangential velocity distribution of subhaloes accreted at $r$, in particular that mentioned in Section \ref{stripped}, we arrive to leading order at the following $\clM$-independent disruption fraction
\beq
f\des(r)= \frac{v\des^2(r)}{v\maxi^2(r)}= \frac{r v\des^2(r)}{GM(r)},
\label{fdes}
\eeq
where $v\des(r)$ is the solution of equation (\ref{vdes}). Note that $f\des(r)$ is independent of the subhalo mass.

\section{RADIAL DISTRIBUTION AND MASS FUNCTION OF STRIPPED SUBHALOES}\label{stripped}

As shown in \citet{SM19}, the virialisation taking place in a major merger yields the memory loss of the system, so the inner properties of the final relaxed object, including those regarding accreted haloes, do not depend on its assembly history. Taking advantage of this important conclusion, in Paper I we concentrated on purely accreting haloes, which notably simplified the calculations without affecting the general validity of the results. Unfortunately, the situation regarding stripped subhaloes is very different. Tidal stripping brakes that possibility because its effects on individual subhaloes are not erased by virialisation, so stripped haloes retain the memory of their past history. As a consequence, to derive the properties of substructure regarding stripped subhaloes we must account for the assembly history of their host haloes. 

In the rest of the present Paper we concentrate on the simplest case of purely accreting haloes or, more exactly, of haloes having been accreting for a long time.\footnote{See Paper I for the relation between the time of the last major merger and the radius from which the object has grown inside-out.} The more complex case of haloes having suffered recent major mergers is addressed in Paper III. 

In these conditions, the mean number of stripped subhaloes per infinitesimal truncated mass and radius within a halo with mass $M\h$ (and virial radius $R\h$) at $t\h$ is
\beqa 
{\cal N}\fin(r,\clM\tr)={\cal
  N}\tr(r,\clM\tr)+\int_{v\des(r)}^{v\maxi(r)}\!\der
v\! \int_{\clM}^{M(r)}\der M~~~\nonumber\\ \times {\cal N}\acc(v,r,M)
\int_{R\tr(v,r,M)}^{R(r,M)} \der r'{{\cal
    N}\fin}_{\rm [M,t(r)]}(r',\clM\tr),~~
\label{corrbis} 
\eeqa
where ${\cal N}\fin_{\rm [M,t(r)]}(r',\clM\tr)$ is the abundance of stripped subsubhaloes at $r'$ inside subhaloes with $M$ accreted at $t(r)$. The first term on the right of equation (\ref{corrbis}), ${\cal N}\tr(r,\clM\tr)$, gives the mean abundance of stripped subhaloes directly arising from the {\it truncation} of subhaloes accreted at $r$ with suited original mass, 
\beqa
{\cal N}\tr(r,\clM\tr)~~~~~~~~~~~~~~~~~~~~~~~~~~~~~~~~~~~~~~~~~~~~~~~~~~~~~~~\nonumber\\
=\int_{v\des(r)}^{v\maxi(r)} \der v\, {\cal
  N}\acc[v,r,\clM(v,r,\clM\tr)]\, \frac{\partial \clM(v,r,\clM\tr)}{\partial
  \clM\tr}.
\label{firstterm0} 
\eeqa
And the second term gives the abundance of stripped subhaloes with $\clM\tr$ that arise from subsubhaloes at $r'$ between the non-truncated radius $R(r,M)$ and truncated one $R\tr(v,r,M)$ of subhaloes with masses $M$ between $\clM$ and $M(r)$ at $t(r)$, ${{\cal N}\fin}_{\rm [M,t(r)]}(r',\clM\tr)$. Note that we take into account that such released subsubhaloes are not further stripped in the host halo (see App.~\ref{App2}). 

As the kinematics of objects with apocentre at $r$ does not depend on their mass (see Sec.~\ref{stripping}), the abundance ${\cal N}\acc(v,r,\clM)$ of accreted subhaloes per infinitesimal original mass, radius and (tangential) velocity in equation (\ref{firstterm}) factorises in the velocity distribution \citep{Jea15} times the mean abundance of accreted subhaloes, ${\cal N}\acc(r,\clM)$, equal to (eq.~[17] of Paper I)
\beq
{\cal N}\acc(r,\clM)= 4\pi\,r^2 \frac{\rho(r)}{M\h}\,{\cal N}\acc(\clM),
\label{ratio}
\eeq
if $\clM<M(r)/3$ and zero otherwise. Thus, taking into account that the mass function of accreted subhaloes ${\cal N}\acc(\clM)$ is very nearly proportional to $\clM^{-2}$ (Paper I), equation (\ref{firstterm0}) leads to the simple expression
\beq
{\cal N}\tr(r,\clM\tr)
= \mu(r)\, {\cal N}\acc(r,\clM\tr),%\equiv \mu(r)\, {\cal N}\acc(r,\clM\tr),
\label{firstterm} 
\eeq
where $\mu(r)$ stands for the average over $v$ from zero (or $v\des(r)$ in case of significant disruption; see Sec.~\ref{destruction}) to $v\maxi(r)$ of the truncated-to-original mass ratio $\clM\tr/\clM$ of subhaloes with original mass $\clM < M(r)/3$ at $r$.\footnote{By ${\cal N}\acc(\clM\tr)$ we mean ${\cal N}\acc$ for a value of the accreted subhalo mass equal to $\clM\tr$.} Note that, ${\cal N}\acc(r,\clM\tr)$ is separable (eq.~[\ref{ratio}]), so is also ${\cal N}\tr(r,\clM\tr)$.

The $\mu(r)$ profile can be readily calculated by monitoring the mass loss through repetitive stripping and heating (eq.~[\ref{last}]) of accreted subhaloes with suited initial properties (see Sec.~\ref{stripping}) and taking the velocity averages for the velocity distribution function given in Appendix \ref{veldis}. The typical concentration of subhaloes accreted by the host halo at $t(r)$ is taken equal to that of haloes with $10^{-2}M(r)$ at that time according to the (time-varying) $M$--$c$ relation found by \citet{Gea08} in simulations similar to those used by SWV and HCFJ. The result is very little sensitive, however, to the particular subhalo mass chosen provided it is between $10^{-1}M(r)$ and $10^{-3}M(r)$.

\begin{figure}
\centerline{\includegraphics[scale=.45,bb= 18 50 560 558]{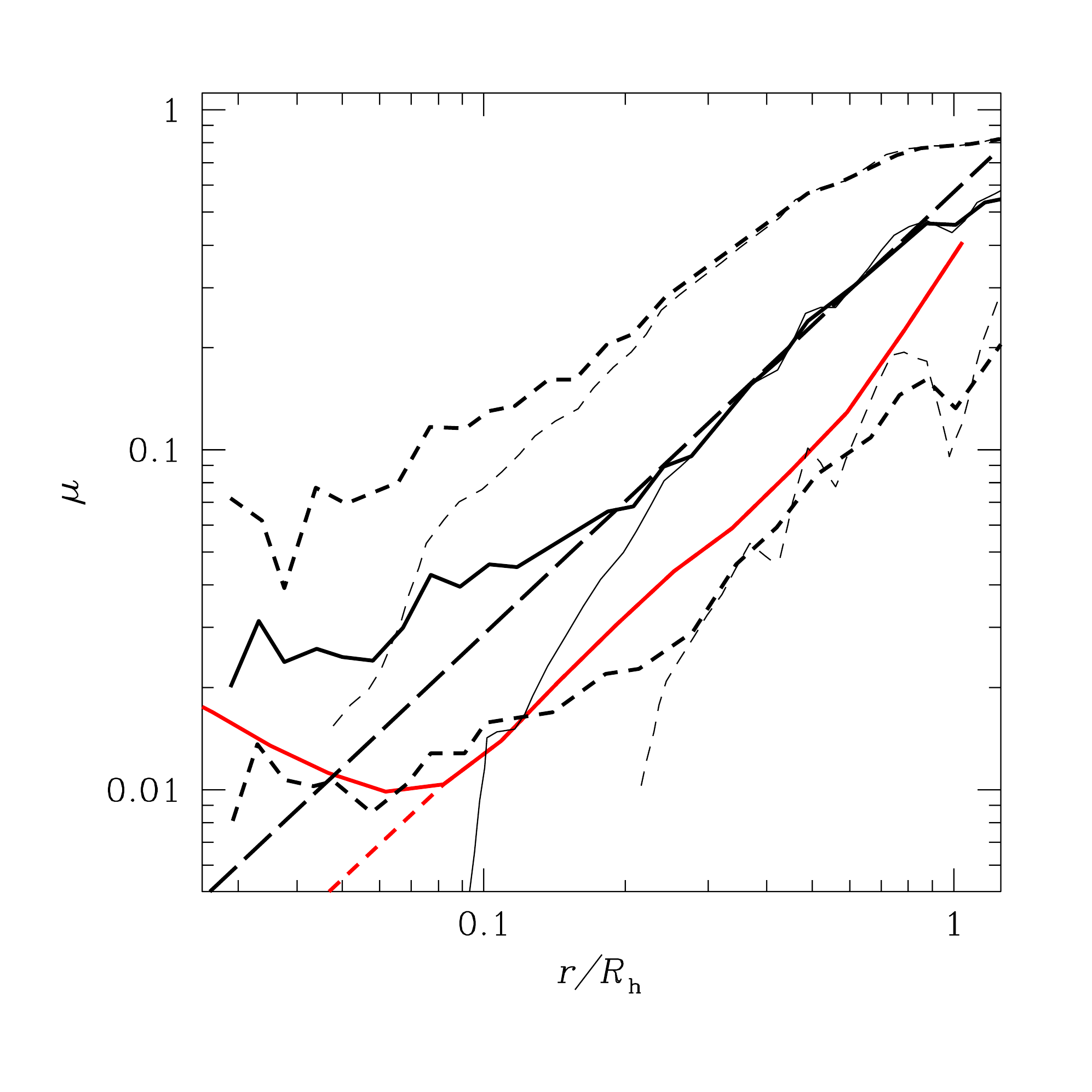}}
\caption{Mean truncated-to-original subhalo mass ratio profile predicted for purely accreting MW-mass haloes (solid red line) for subhaloes with concentrations equal to that of haloes with $10^{-2}M(r)$ at $t(r)$ according to the empirical $M$--$c$ relation by \citet{Gea08}. For comparison we plot the empirical {\it median} mass ratio profile of resolved (thick black lines) and resolved plus orphan (thin black lines) subhaloes found by HCFJ in the Level 1 Aquarius halo A (thin and thick dashed black lines give the corresponding upper/lower $1 \sigma$ percentiles, respectively), which has been accreting since $r=0.08R\h$. We also plot the power-law {\it median} profile of index 1.3 used in the HCFJ model (long-dashed black line) and the power-law  extrapolation of our theoretical mean profile (dashed red line) at the radius $r<0.08R\h)$, from which the halo A has evolved by accretion as considered in our model.} 
(A colour version of this Figure is available in the online journal.)
\label{f2}
\end{figure}

The resulting $\mu(r)$ profile, which coincides down to one thousandth $R\h$ for the cases of null or moderate disruption, is shown in Figure \ref{f2}. For comparison we also plot the results found by HCFJ for the MW-mass halo A in the Level 1 (maximum resolution) Aquarius simulation (SWV). Note that this halo is particularly well-suited to the comparison with the predictions of CUSP for purely accreting haloes because it has been accreting since $z \sim 6$, i.e. it has been growing inside-out from $r/R\h\sim 0.08$ (see Fig.~1 of Paper I). Thus, the comparison is only meaningful down to that radius. To see how sensitive the results are on the modelling below $r=0.08R\h$ we also consider the theoretical $\mu(r)$ profile extrapolated by a simple power-law to small radii. Of course, we cannot pretend that the predicted $\mu(r)$ profile coincides with that found by HCFJ in the A halo: ours refers to the {\it mean} truncated-to-original subhalo mass ratio, while HCFJ's refers to the {\it median} one. More importantly, as mentioned in Paper I, the $\mu$ profile derived by HCFJ is for subhaloes of all levels while ours holds for first-level subhaloes only. Since these latter are the only subhaloes being stripped, including all level subhaloes should notably enhanced the resulting mass ratio. This would explain that our mean $\mu$ profile is substantially lower than the median $\mu$ profile found by HCFJ, contrarily to what would be expected for a lognormal distribution of mass ratios referring to the same subhalo population. Note also that the validity of our mean profile is supported by the fact that, as we will see, it leads to the right amplitude of the predicted MF of stripped subhaloes. On the contrary, were the predicted mean $\mu(r)$ profile raised to at least the position of the empirical median profile, the resulting MF would be a factor 2 higher. In Figure \ref{f2} we also plot a power-law form with index $1.3$, proportional to the median truncated-to-original mass ratio profile in the HCJF model. 

\begin{figure}
\centerline{\includegraphics[scale=.45,bb= 18 50 560 558]{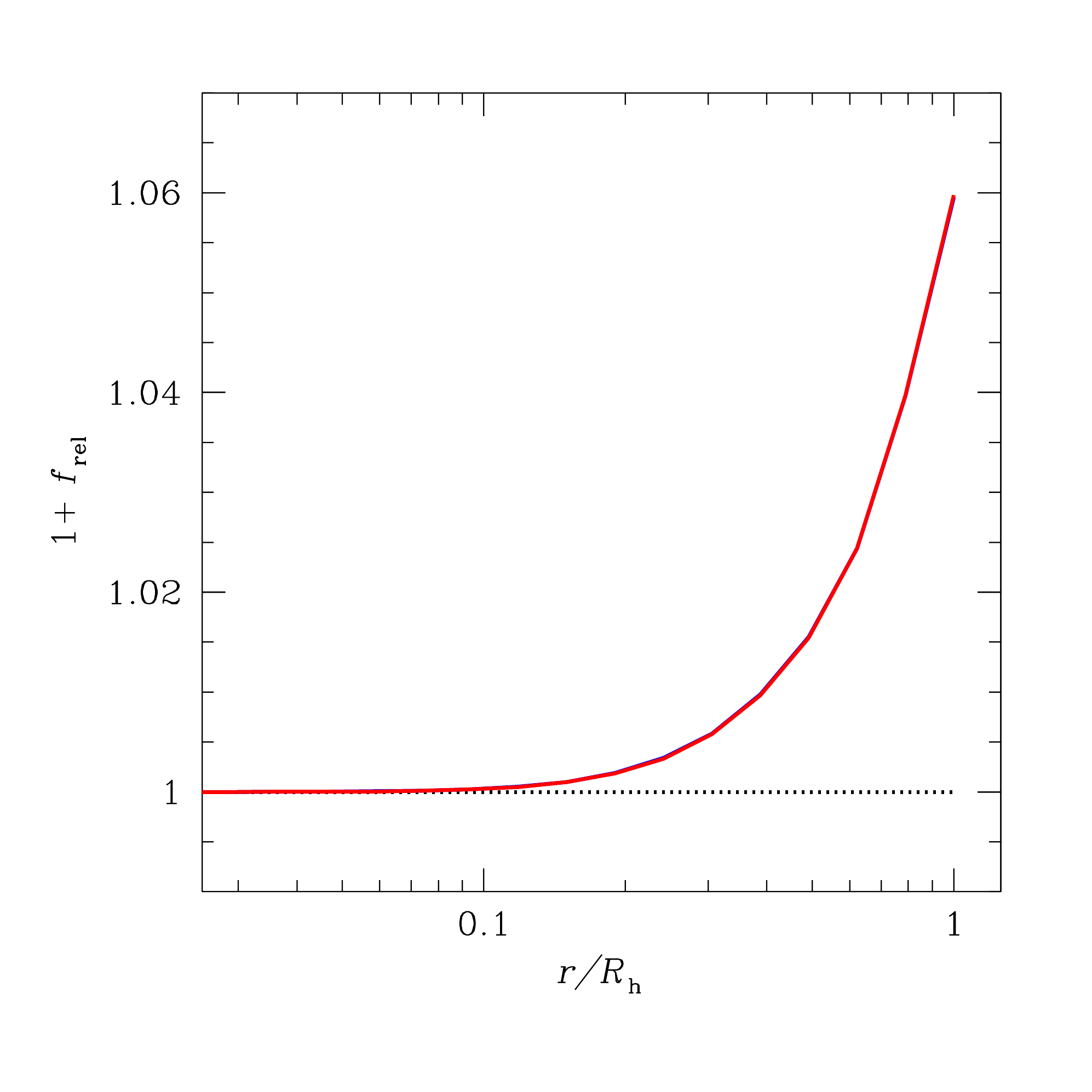}}
\caption{Function $f\trb(r)$ predicted for MW-mass haloes. At the scale of the Figure, the solutions corresponding to a SWV-like simulation and a real 100 GeV WIMP universe overlap.}
(A colour version of this Figure is available in the online journal.)
\label{f3}
\end{figure}

Once the $\mu(r)$ profile has been determined, defining $1+f\trb(r,\clM\tr)$ as ${\cal N}\fin(r,\clM\tr)/{\cal N}\tr(r,\clM\tr)$, equation (\ref{corrbis}) can be rewritten in the form
\beq
{\cal N}\fin(r,\clM\tr)=
[1+f\trb(r)]\,\mu(r)\,{\cal N}\acc(r,\clM\tr),
\label{M} 
\eeq
where $f\trb(r)$ is the proportion of stripped subhaloes previously locked as subsubhaloes within accreted subhaloes per each stripped subhalo directly arising from the truncation of an accreted subhalo. That proportion is the solution of the Fredholm integral equation of second kind that follows from equation (\ref{corrbis}) (see App.~\ref{App1}), 
\beqa
1+f\trb(r)=1+3\,\frac{r^3}{R\h^3}\,\int_0^{R\h} \der r'\,r'^2\,F(r',r)\,~~~~~~~~~~~~~~~~~\nonumber\\
\times\, [1+f\trb(r')]\frac{\mu(r')}{\mu(r)}\,\frac{\rho(r')}{\bar \rho(R\h)},~~~~
\label{final3}
\eeqa
where $F(r',r)$ is the cumulative velocity distribution function (see App.~\ref{veldis} for the corresponding differential form) for the velocity $v$ such that $Q_{\nu(v,r)}(v,r)=r'$, where $Q_{\nu(v,r)}(v,r)$ is the ratio of truncated-to-original subhalo radii after $\nu(v,r)$ passages by pericentre calculated in Section \ref{stripping}. As shown in Figure \ref{f3}, the fraction $f\trb(r)$ increases with increasing radius and reaches a maximum value of 0.06 in current haloes. We remark that, even though the contribution of subsubhaloes to the properties of substructure has been included in previous analytic models (e.g. \citealt{TB04,Zeabis05}) as well as simulations (e.g. \citealt{Hea12,Hea18}), a quantitative estimate of the contribution of released subsubhaloes as that given here was missing.

Note that, $f\trb(r)$ depends on disruption through $F(r',r)$. Yet, like in the case of $\mu(r)$, we have found no significant difference down to one thousandth $R\h$ between null and moderate disruption. Thus, disruption does not affect the predicted abundance of stripped subhaloes, ${\cal N}\fin(r,\clM\tr)$, which supports the claim by \citet{vdB18} that it is likely a numerical effect (see also \citealt{EP20}).

Equation (\ref{M}) states that the abundance of stripped subhaloes is proportional to the abundance of accreted subhaloes with proportionality factor equal to $[1+f\trb(r)]\mu(r)$. This resembles the relation found in the HCFJ model with the median truncated-to-original subhalo mass ratio playing the role of $\mu(r)$. We remark, however, that the relation with the mean truncated-to-original subhalo mass ratio profile is essentially exact (see eq.~[\ref{firstterm}]; but see Sec.~\ref{masses}), while the equivalent one in the HCFJ model is a relation between typical quantities holding for subhaloes.

\begin{figure}
\centerline{\includegraphics[scale=.45,bb= 18 50 560 558]{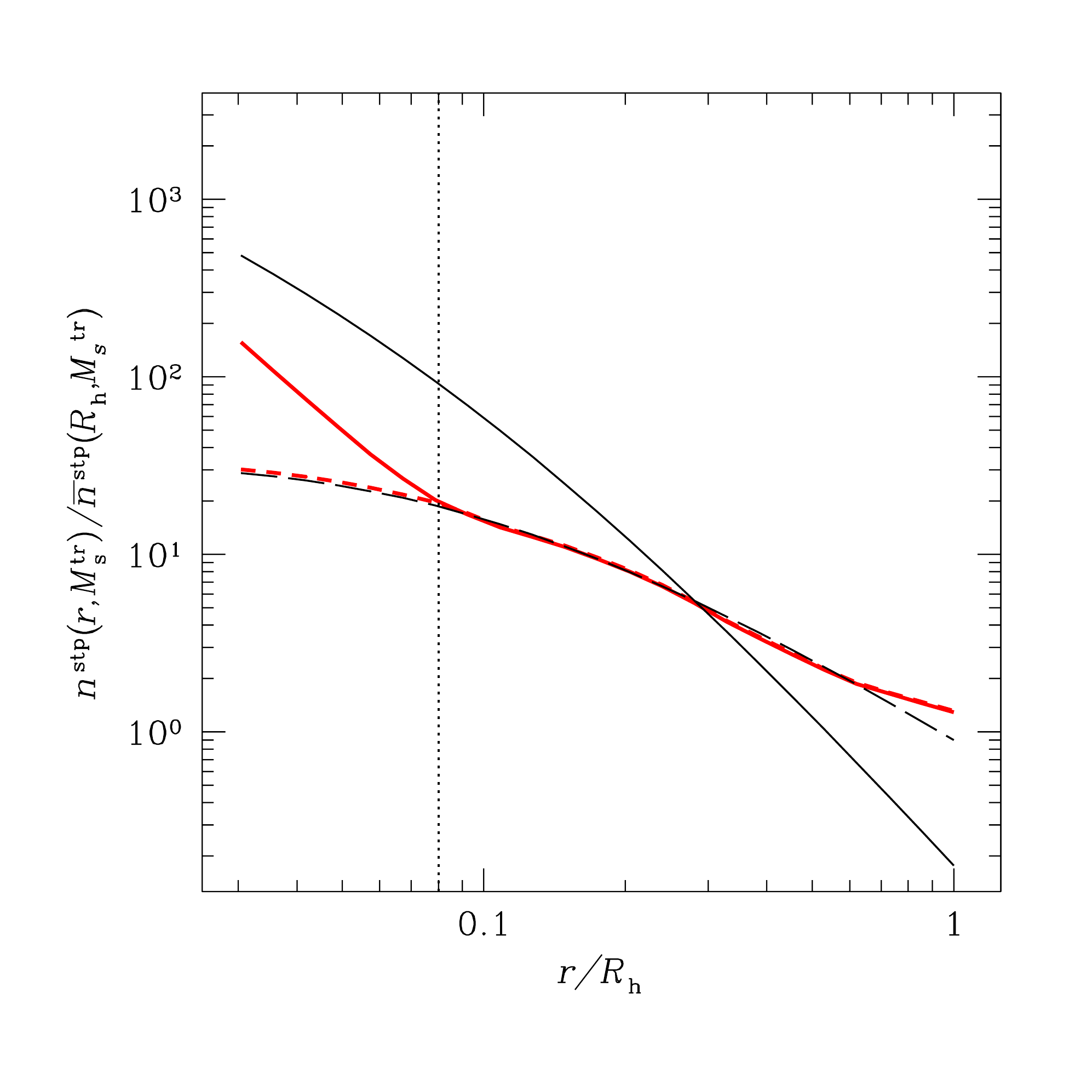}}
\caption{Scaled number density profiles for stripped subhaloes of different masses and similar concentrations predicted for purely accreting MW-mass haloes from the two versions of the $\mu(r)$ profiles obtained using the \citet{Gea08} $M$--$c$ relation shown in Figure \ref{f3} (same lines). For comparison we plot the profile of the form $\propto r^{1.3}\rho(r)$ (long-dashed black line) providing a good fit to the scaled subhalo density profile of the Level 1 Aquarius halo A (HCFJ) and the scaled mass density profile $\rho(r)$ of that halo (solid black line). The vertical dotted line marks the radius out of which the halo A has evolved by accretion as considered in CUSP.}
(A colour version of this Figure is available in the online journal.)
\label{f4}
\end{figure}

From equations (\ref{M}) and (\ref{ratio}) we see that ${\cal N}\fin(r,\clM\tr)$ is separable like ${\cal N}\acc(r,\clM\tr)$. This is the reason that the mean number density profile per infinitesimal mass of stripped subhaloes, $n\fin(r,\clM\tr)\equiv {\cal N}\fin(r,\clM\tr)/(4\pi r^2)$, scaled to the mean number density per infinitesimal mass of such subhaloes, $\bar n\fin(R\h,\clM\tr)\equiv 3{\cal N}\fin(\clM\tr)/(4\pi R\h^3)$, takes the mass-independent form (see eqs.~[\ref{ratio}]--[\ref{M}])
\beq
\frac{n\fin(r,\clM\tr)}{\bar n\fin(R\h,\clM\tr)}\!=\!
\frac{[1+f\trb(r)]\mu(r)}{\overline {(1+f\trb)\mu}(R\h)} \frac{\rho(r)}{\bar\rho(R\h)}.
\label{new}
\eeq
Thus the scaled number density profiles of stripped subhaloes of different truncated masses $\clM\tr$ overlap in one single profile (but see Sec.~\ref{masses}), as found in Paper I for accreted subahlos and in agreement with the results of SWV (see also \citealt{Lea09}). However, contrarily to what happens with the profile of accreted subhaloes that of stripped subhaloes is not proportional to $\rho(r)$ but bends downwards at small radii (Fig.~\ref{f4}) also in agreement with the results of simulations (\citealt{Gea98,Dea04,Gea04,NK05,Dea07}; SWV). Equation (\ref{M}) shows that such a bending is due to the factor $\mu(r)$, which, as we will see in Section \ref{dDM}, entails an increasing abundance of dDM towards the halo centre. More importantly, the predicted profile recovers that found in the Aquarius halo A regardless of the exact form of $\mu(r)$ below $r=0.08R\h$, showing that the agreement between our prediction for purely accreting haloes and the properties of the halo A is compelling regardless of the growth of the halo before $z=6$. We thus find the same result as the HCFJ model using the median truncated-to-original subhalo mass ratio profile instead of the mean one. This could be foreseen since any constant shift between the two profiles cancels when deriving the `scaled' density profile.

\begin{figure}
\centerline{\includegraphics[scale=.45,bb= 18 40 560 570]{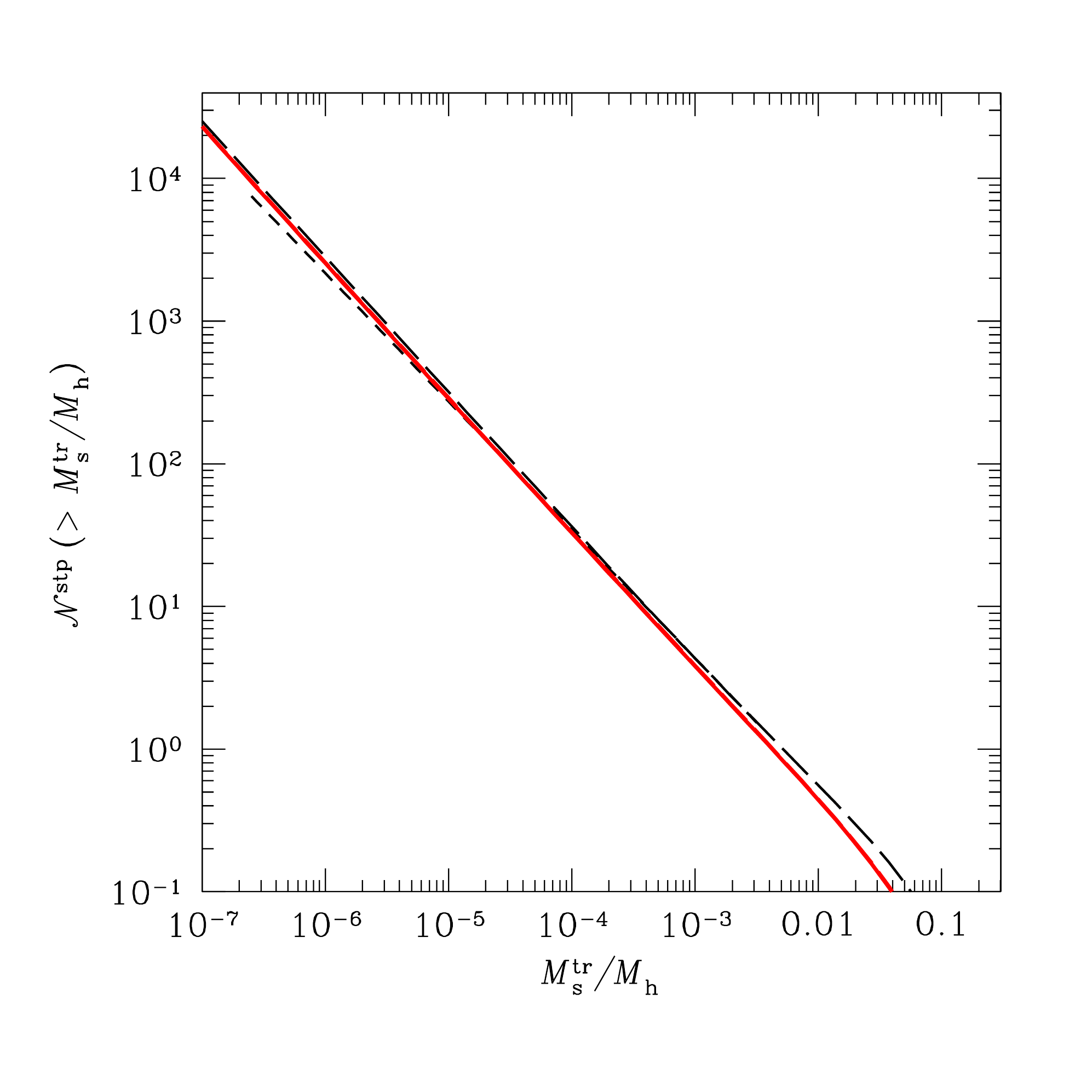}}
\caption{Cumulative MF of stripped subhaloes
  predicted in purely accreting MW-mass haloes from the mean $mu(r)$ profile (solid red line) and its extrapolation to $r<0.08R\h$ (dashed red line) compared to the MF of the Level 1 Aquarius halo A obtained by SWV (short-dashed black line) and the general MF for haloes of that mass found by \citet{Hea18} (long-dashed black line).}
  (A colour version of this Figure is available in the online journal.)
\label{f5}
\end{figure}

Integrating over $r$ the abundance of stripped subhaloes given in equation (\ref{M}), we obtain the differential MF of stripped subhaloes per infinitesinal mass,
\beq
{\cal N}\fin(\clM\tr)=\overline {(1+f\trb)\mu}(R\h)\,{\cal N}\acc(\clM\tr)\,.
\label{MF}
\eeq
According to this relation, ${\cal N}\fin(>\! \clM\tr)$ is proportional to ${\cal N}\acc(>\! \clM\tr)$, which agrees with the results of simulations (HFCJ) at least at intermediate and low masses (see below). As shown in Figure \ref{f5}, the corresponding cumulative MF is in fairly good agreement with that found in simulations. Specifically, at low-masses it is approximately a power-law form with a logarithmic slope of $\sim -0.95$ (the logarithmic slope of the MF of accreted subhaloes varies from $-0.94$ to $-0.97$ from high to low subhalo masses; Paper I), which is intermediate between the slopes of $-0.9$ and $-1$ reported by SWV and \citet{Dea07}, respectively, and very close to $-0.94$ \citep{BK10,Gea11} and $-0.95$ at $\clM/M\h\sim 10^{-5}$ (HCFJ; \citealt{Hea18}). 

\section{Abundance and Radial Distribution
of Diffuse Dark Matter}\label{dDM}

\begin{figure}
\centerline{\includegraphics[scale=.56,bb= 30 70 500 495]{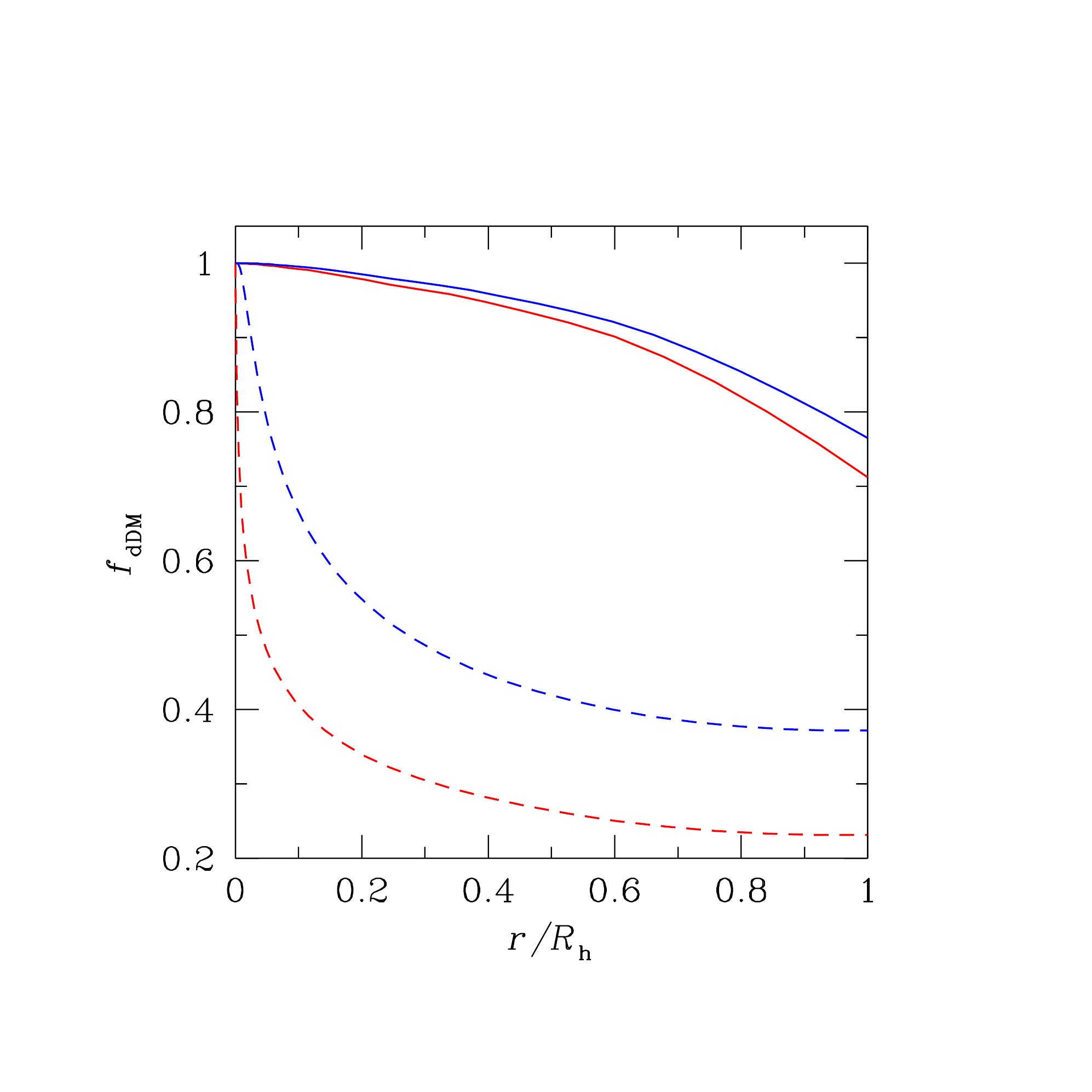}}
\caption{Total (accreted plus stripped) dDM mass fraction profiles in purely accreting MW-mass haloes predicted in a real 100 GeV WIMP universe (solid red line) and a SWV-like simulations (solid blue line). Dashed lines give the corresponding accreted dDM mass fraction profiles derived in Paper I.}
(A colour version of this Figure is available in the online journal.)
\label{f6}
\end{figure}

The total mass density of the halo at $r$ results from the contribution of stripped subhaloes and dDM. We thus have that the total dDM mass fraction $f\dc\fin(r)$ at $r$ satisfies
\beq
[1-f\dc\fin(r)]\rho(r)\!=\!\frac{1}{4\pi r^2}\int_{M\res}^{M\h}\!\! \der \clM\tr\, \clM\tr\,{\cal N}\fin(r,\clM\tr),
\label{rhod}
\eeq
where $M\res$ is the minimum halo mass at the origin of dDM (see Paper I). Taking into account equations (\ref{M}) and (\ref{ratio}), equation (\ref{rhod}) leads to
\beq
1-f\dc\fin(r)=[1+f\trb(r)]\,\mu(r)[1-f\dc\acc(r)].
\label{fdDM}
\eeq

In Figure \ref{f6} we plot the total dDM mass fraction profile $f\dc\fin(r)$ for purely accreting MW-mass haloes (eq.~[\ref{fdDM}]) for the same illustrative cases as used in Paper I to calculate the accreted dDM mass fraction at $r$, $f\dc\acc(r)$ (also plotted in Figs.~\ref{f6}): 1) a real 100 GeV WIMP universe with minimum halo mass $M\res= 10^{-6}$ \modot and $t\res$ equal to the time of decoupling and 2) a SWV-like simulation starting at $t\res=0.0124$ Gyr ($z\res=127$) and with a resolution mass of $M\res=4.4 \times 10^5$ \modotc. As can be seen, the dDM mass fraction in current haloes is quite large; it is sill $\sim 0.75$ at $R\h$ where it reaches the minimum value. Moreover, a substantial fraction has been stripped from subhaloes. Specifically, in the 100 GeV WIMP universe, we find that 92 \% of the total mass of MW-like haloes is in the form of dDM, with 33 \% directly accreted from the intra-halo medium (Paper I; see also \citet{AW10}). And in SWV-like simulations, 95 \% of the total mass of those haloes is typically in the form of dDM, with 51\% directly accreted (Paper I; see also \citet{Wa11}).

The relation (\ref{fdDM}) allows one to rewrite the scaled subhalo number density of stripped subhaloes (\ref{new}) in the form 
\beq
\frac{n\fin(r,\clM\tr)}{\bar n\fin(R\h,\clM\tr)}=\frac{\clf(r)}{\bar \clf(R\h)}\,\frac{\rho(r)}{\bar\rho(R\h)}, 
\label{fin}
\eeq
where $\clf(r)$ is the stripped to accreted subhalo mass ratio profile, equal to $[1-f\fin\dc(r)]/[1-f\acc\dc(r)]$. Comparing expressions (\ref{fin}) and (\ref{new}), we see that the bending of the scaled number density profile of stripped subhaloes with respect to the scaled density profile of the host halo found in simulations, shown to obey the factor $[1+f\trb(r)]\mu(r)$, can also be seen to obey the factor $[1-f\dc\fin(r)]/[1-f\dc\acc(r)]$. Thus, such a bending is due to the increasing abundance of dDM mass towards the halo centre due to the stripping of subhaloes (SWV; \citealt{Aea09,Gea12}; \citealt{Hell16,Fiea20}). Likewise, taking into account the relation (\ref{fdDM}), the MF of stripped subhaloes, equation (\ref{MF}), can be rewritten in the form
\beq
{\cal N}\fin(\clM\tr)=\bar \clf(R\h)\,{\cal N}\acc(\clM\tr).
\label{MF1}
\eeq

\section{The Effect of the Mass-Dependent Concentrations}\label{masses}

In Sections \ref{stripped} and \ref{dDM} we have taken into account that subhaloes accreted at different times $t(r)$ have different typical concentrations. But halo concentration also slightly depends on their mass. In this section we explicitly account for this latter dependence. 

A first important consequence of such a dependence is that, according to the results of Section \ref{stripping}, the HCJF condition 3 would be only approximate. This is not contradictory with the results of simulations. That condition, which followed from the HCFJ stripping model where subhaloes were approximated by isothermal spheres with mass-independent density profiles, was shown by these authors to be consistent with the results of simulations but not confirmed by them. Certainly it was shown to lead to a mass-independent scaled number density profile of stripped subhaloes as found in simulations (e.g. SWV; \citealt{Lea09}). But this latter result has also been shown to be only approximate. The more detailed study carried by \citealt{Hea18} has found, indeed, that the scaled number density profile of subhaloes actually depends on their mass (see the discussion below). On the other hand, as shown next, our stripping model in its more accurate version taking into account the mass dependence of halo concentration also leads to the same approximate result, so we must not worry about the idea that the HCFJ condition 3 is approximate just as our previous results. 

Having said this, we can follow essentially the same steps as in Section \ref{stripped} using the full time and mass dependence encoded in the $M$--$c$ relation. Of course, if we are to recover the radial abundance and MF of stripped subhaloes observed in simulated haloes, we should use the empirical $M$--$c$ relation found e.g. by \citep{Gea08} affected by the limited mass resolution of simulations giving rise to those empirical properties as done in Section \ref{stripped}. But we are also interested in deriving the real (unbiased) radial abundance and MF of stripped subhaloes predicted by CUSP, so we will use, in addition, the $M$--$c$ relation directly arising from our formalism (see \citealt{Jea19} in preparation) and see the differences in the results obtained from both $M$--$c$ relations.

\begin{figure}
\centerline{\includegraphics[scale=.45,bb= 18 50 560 558]{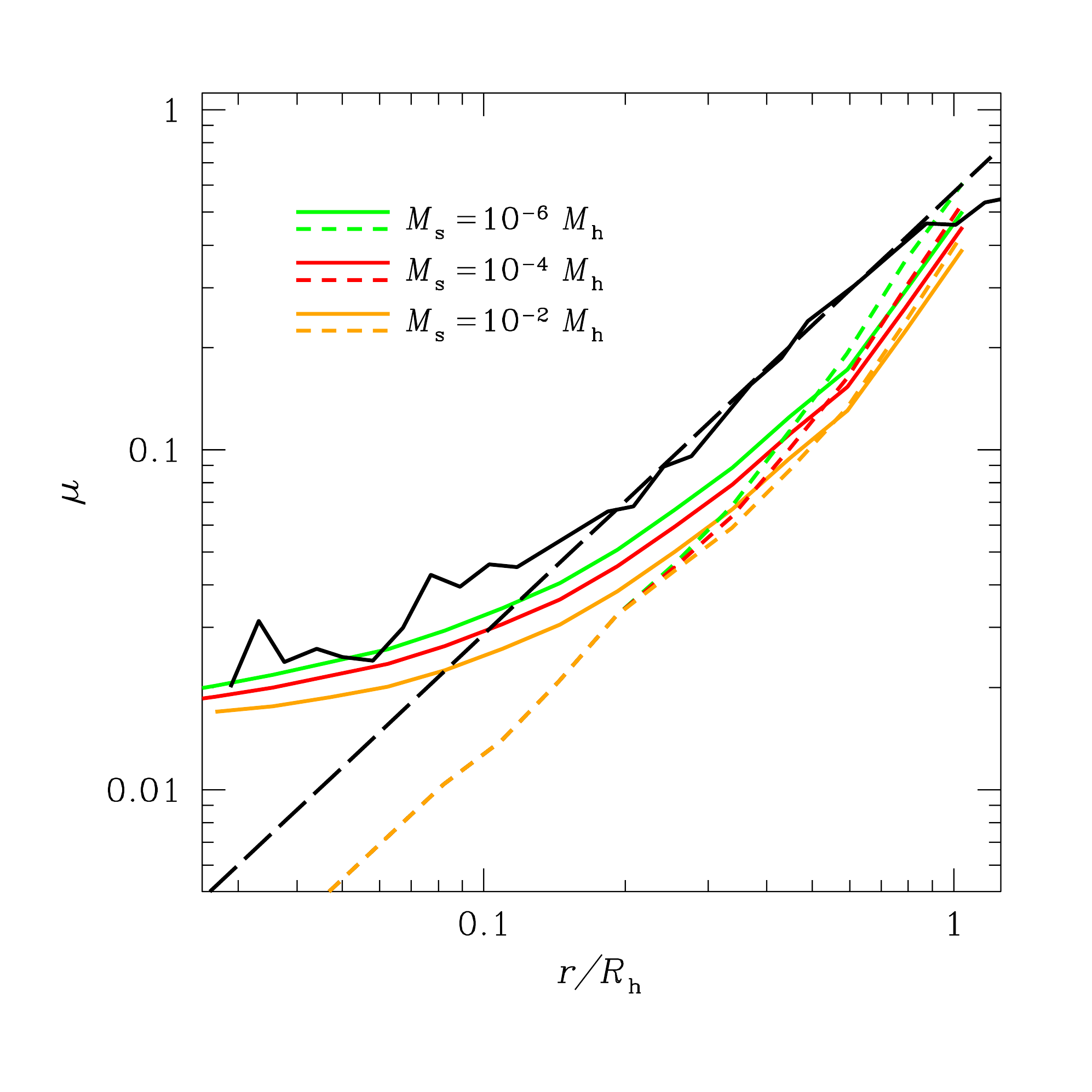}}
\caption{Mean truncated-to-original subhalo mass ratio profile predicted in purely accreting MW-mass haloes for different subhalo masses (coloured lines) from the respective $\mu(r,\clM)$ profiles obtained using the CUSP $M$--$c$ relation (solid lines) and the extrapolated $\mu(r,\clM)$ profiles obtained from the \citet{Gea08} $M$--$c$ relation (dashed lines). As references, we also plot the empirical median truncated-to-original subhalo mass ratio profile found by HCFJ in the Level 1 Aquarius halo A (solid black line) and its modelled version (long-dahsed black line), both depicted in Figure \ref{f4}.}
(A colour version of this Figure is available in the online journal.)
\label{f7}
\end{figure}

In Figure \ref{f7} we show the $\mu$ profiles obtained for several subhalo masses. At large $r$ the solutions arising from the two $M$--$c$ relations are very similar. This is not unsurprising since both relations behave very similarly at low-$z$. Specifically, the solutions obtained for different subhalo masses are more or less shifted vertically as expected, though the amplitude of those shifts slightly depend on the particular $M$--$c$ relation used due to their slightly different shape.\footnote{At $z=0$, Gao et al. $M$--$c$ relation is linear in log-log, while that predicted by CUSP becomes slight shallower towards small masses.} At $r\la 0.2R\h$, however, the solutions drawn from the two $M$--$c$ relations deviate from each other. In the case of the empirical relation, the (extrapolated) curves for different masses converge at $r<0.02R\h$ and are kept with the same power-law shape as at larger radii. (The reason for their convergence is that the log-log $M$--$c$ relation provided by Gao et al. becomes an horizontal line at $z$ somewhat higher than 3, meaning that all subhaloes with different masses have identical concentrations there.) While, in the case of the theoretical $M$--$c$ relation, the curves for the different subhalo masses show the same vertical shifts as at large $r$ but deviate from a power-law of index $\sim 1.3$ and become increasingly less steep towards the centre. 

A good approximation for those $\mu$ profiles, particularly for the case of the CUSP $M$--$c$ relation, is thus
\beq
\mu_{[M\h,t\h]}(r,\clM)\approx \mu_{[M\h,t\h]}(r,\Mz)\,G\left(\clM\right)
\label{newmu}
\eeq
with $\Mz$ an arbitrary mass and $G(\clM)$ equal to $(\clM/\Mz)^\alpha$ with $\alpha\approx-0.03$ in the case of the CUSP $M$--$c$ relation and the same expression at $r\ge 0.2R\h$ (otherwise with no dependence on $\clM$) in the case of Gao et al.'s $M$--$c$ relation. Note that the concentration of the host halo at $t(r)$ is not the typical concentration of haloes at that time but $r/r_s$ so that decreasing $\clM$ by some factor and keeping $M\h$ fixed is not equivalent to keeping $\clM$ fixed and increasing $M\h$ by the same factor. Consequently, $\mu_{[M\h,t\h]}(r,\clM)$ is not a power-law of $M\h$ with index $-\alpha$.

Then, the same derivation leading from equation (\ref{firstterm0}) to equation (\ref{firstterm}) leads to
\beqa
{\cal N}\fin_{[M\h,t\h]}(r,\clM\tr)
= g(\alpha)[1+f\trb(r,\clM\tr)]\mu_{[M\h,t\h]}(r,\clM\tr)\nonumber\\\times{\cal N}\acc_{[M\h,t\h]}(r,\clM\tr),~~~~~~~~~~~~~~~~~~
\label{newfirstterm} 
\eeqa
with the factor $g(\alpha)$ equal to 1 or $1-\alpha$ depending on whether $G(\clM)$ in equation (\ref{newmu}) is unity or not. Note that the function $f\trb$ solution of the differential equation (\ref{final3}) for the function $\mu_{[M\h,t\h]}(r,\clM\tr)$ also depends now on $\clM\tr$. 

\begin{figure}
\centerline{\includegraphics[scale=.45,bb= 18 50 560 558]{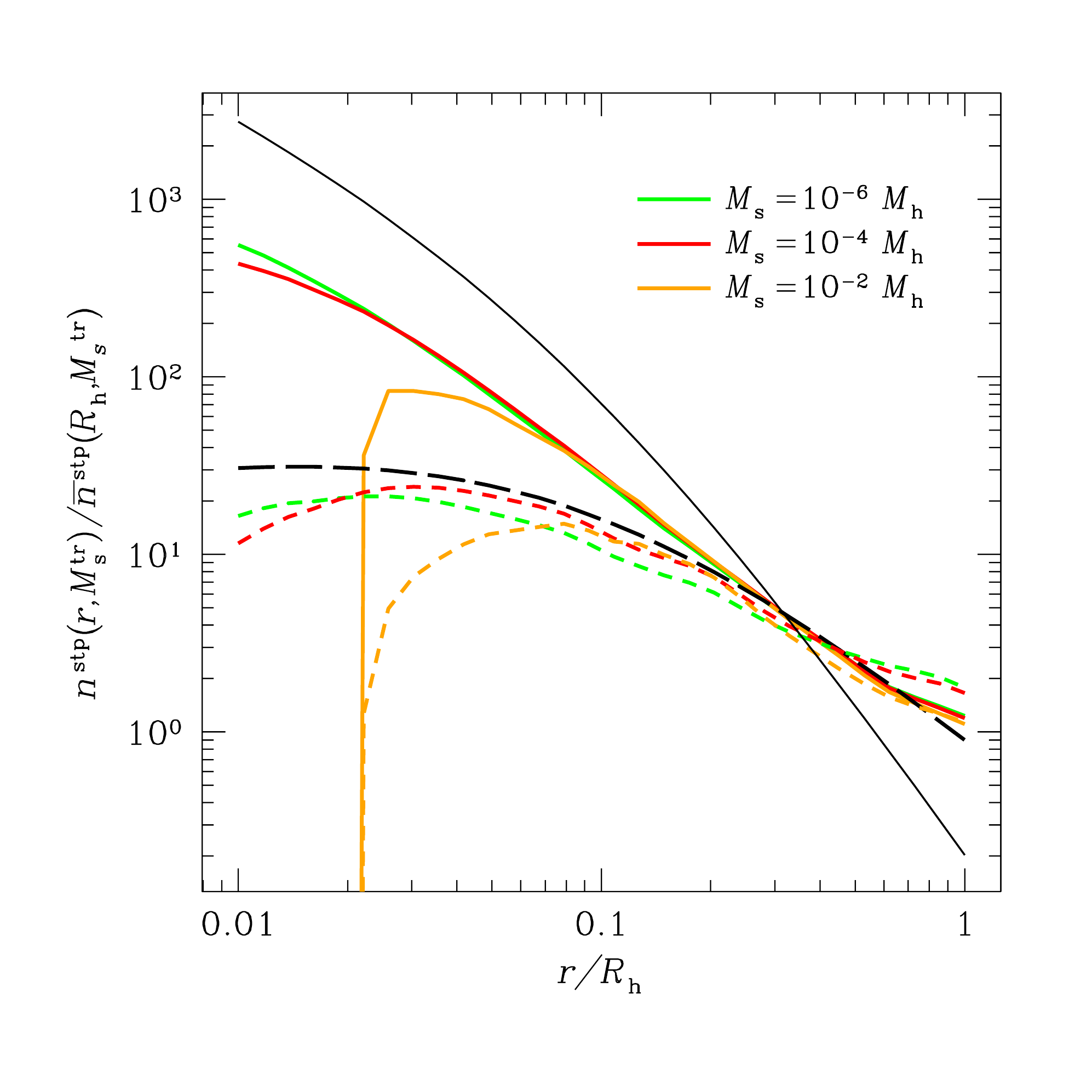}}
\caption{Scaled number density profiles of stripped subhaloes of different masses (coloured lines) predicted for purely accreting haloes using the $M$--$c$ relation found in simulations by \citet{Gea08} (dashed lines) and the theoretical $M$--$c$ relation predicted by CUSP (solid lines). The long-dashed black line is the HCFJ fit to the profile found for low enough mass subhaloes in the Level 1 Aquarius halo A. The solid black line is the scaled halo density profile. All empirical data have been converted to $M\h$.}
(A colour version of this Figure is available in the online journal.)
\label{f8}
\end{figure}

The resulting scaled number density profiles of stripped subhaloes of different masses obtained from direct calculation not using the approximation (\ref{newmu}) are shown in Figure \ref{f8}. In the case of the empirical $M$--$c$ relation they are relatively close to the empirical profile of the Aquarius halo A found by SWV and HCFJ, except for the fact that they do not exactly overlap with each other owing to the fact that the vertical shifts in the corresponding $\mu$ profiles are not constant over all radii. On the contrary, the constant vertical shifts found in the case of the theoretical $M$--$c$ relation go unnoticed in the scaled number density profiles of stripped subhaloes of different masses, which thus overlap. Strictly speaking the profiles for different $\clM$ show different cutoffs, which simply reflects that no subhalo with a given mass $\clM$ can be accreted by the halo of mass $M(r)< 3\clM$ as this would cause a major merger with the destruction of the subhalo (see Paper I). However, the resulting number density profile is in this case less bent with respect to the $\rho(r)$ profile at small radii than in the Aquarius halo A. This disagreement does not mean, of course, that the theoretical prediction is wrong. Rather the contrary, what would be biased is the empirical profile due to the limited mass resolution of simulations affecting the $M$--$c$ relation. Our prediction suffer, instead, from the fact that, in its current version, our stripping model does not include the effects of dynamical friction which are particularly strong at the subhalo high-mass end (see below).

Even though the scaled number density of subhaloes of different masses nearly overlap, the non-scaled ones do not because of the different total number of subhaloes of different masses. In fact, the detailed, non-scaled as well as scaled, number density profiles of massive stripped subhaloes are found to be cuspier than those of less massive subhaloes \citep{Hea18}. As discussed by these authors, this result seems to be the consequence of dynamical friction, which causes massive subhaloes to migrate towards the centre of the host halo. In fact, dynamical friction would not only alter the `natural' density profiles of stripped subhaloes but also their MF. Indeed, even though the MF does not depend on the radial location of subhaloes, it appears that dynamical friction also affects the stripping itself of massive subhaloes and, hence, the MF of stripped subhaloes, too. 

\begin{figure}
\centerline{\includegraphics[scale=.45,bb= 18 50 560 555]{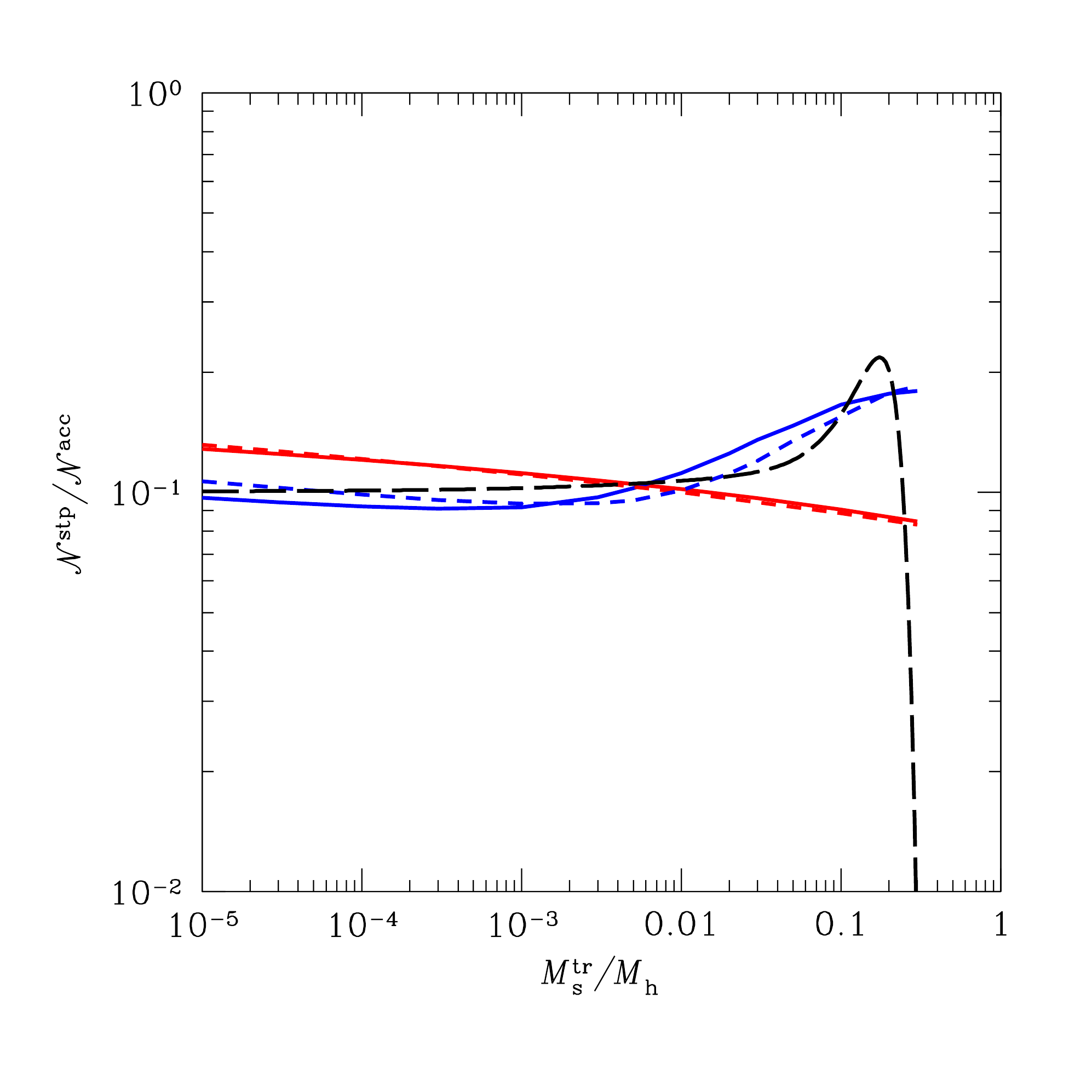}}
\caption{Ratio of the differential MFs of stripped and accreted subhaloes predicted for haloes of different masses (with the appropriate scaling; see text) obtained using the \citet{Gea08} and CUSP $M$--$c$ relations (dashed and solid lines, respectively) compared to the ratio found in simulations by \citet{Hea18} (long-dashed black line), both of them scaled with halo mass as explained in the text. The two solutions depicted are those obtained from the stripping model of Section \ref{stripping} but accounting for the mass-dependent subhalo concentration (red lines) and the same model with suppression of stripping when the mass of subhaloes is larger than $M(r\per)$ (blue lines).}
(A colour version of this Figure is available in the online journal.)
\label{f9}
\end{figure}

In Figure \ref{f9} we depict the ratio between the differential MFs of stripped and accreted subhaloes as function of the scaled mass $m=\clM\tr/M\h$ that result from integration over $r$ of the relation (\ref{newfirstterm}) with the $\mu$ profiles obtained from the two different $M$--$c$ relations. The ratios for different halo masses have been multiplied by $\bar\mu_{[\Mhz,t\h]}(\Rhz,\Mz)/\bar\mu_{[M\h,t\h]}(R\h,\Mz)$ where $\Mhz$ is an arbitrary halo mass and $\Rhz$ its corresponding virial radius so that the curves for different halo masses overlap (see eqs.~[\ref{newmu}] and [\ref{newfirstterm}], with $f\trb$ neglected in front of unity). That scaling factor turns out to behave as $(M\h/\Mhz)^\eta$ with  $\eta=-0.09$ and $-0.08$ in the Gao et al. and CUSP $M$--$c$ relations, respectively. For comparison, we also plot the ratio found for different halo masses by \citep{Hea18} multiplied by the factor $(M\h/\Mhz)^{-0.1}$, which also causes those empirical curves to overlap \citep{Hea18,Rea16}. Note that the empirical MF of stripped subhaloes plotted in this Figure is not normalised as the original MF in \citet{Hea18}. The reason for this is that the latter holds for subhaloes of all levels, while we are interested in the MF of first-level subhaloes only. We have thus shifted it, downwards keeping the same proportion with respect to the MF of accreted subhaloes correctly renormalised so as to hold for first-level subhaloes only (see Paper I). Even though, that empirical MF of stripped subhaloes is more comparable to the predicted MF, we cannot guarantee yet that its normalisation is fully correct. 

The theoretical ratios derived from both $M$--$c$ relations substantially deviate from the empirical one. While the former decreases slowly with increasing $m$, the latter increases, shows a marked bump, and rapidly falls to zero at the high-mass end \citep{Hea18,Ji16}. This disagreement is due to the fact that our stripping model fails for massive subhaloes. Indeed, as pointed out by \citet{Hea18}, the tidal force from the host halo on very massive subhaloes becomes less important than its self-gravity, so stripping is less effective. Properly accounting for that effect is out of scope of this Paper because our model does not include dynamical friction, while that effect alters the pericentre of subhaloes and, hence, the mass of the host halo seen by subhaloes there, $M[r\per(r,v)]$. To see the kind of effect this condition may have we depict in Figure \ref{f9} the result of suppressing stripping when the geometrical mean of $\clM$ and $\clM\tr(v,r,\clM)$ is more massive than $M[r\per(r,v)]$.\footnote{That geometrical mean increases the mass of the stripped subhalo so as to compensate that the real $r\per$ is smaller than calculated without dynamical friction.} As can be seen, the trend of the new MF (normalised as the empirical one) greatly improves, indeed. 

But this effect increases the ratio towards high masses, which goes in the opposite direction from that needed to find the exponential cutoff at the high-mass end of the empirical ratio. That cutoff is likely due, once again, to the effects of dynamical friction. Indeed, extremely massive subhaloes rapidly migrate to the halo centre and merge with the central subhalo, so they `disappear' rather than stay a little less stripped. Of course, even without that effect, the predicted differential MFs of stripped subhaloes ends up by falling off to zero as $\clM\tr$ approaches $M\h/3$ due to the similar cutoff present in the differential MF of accreted subhaloes (see Paper I).

\begin{figure}
\centerline{\includegraphics[scale=.45,bb= 18 46 550 570]{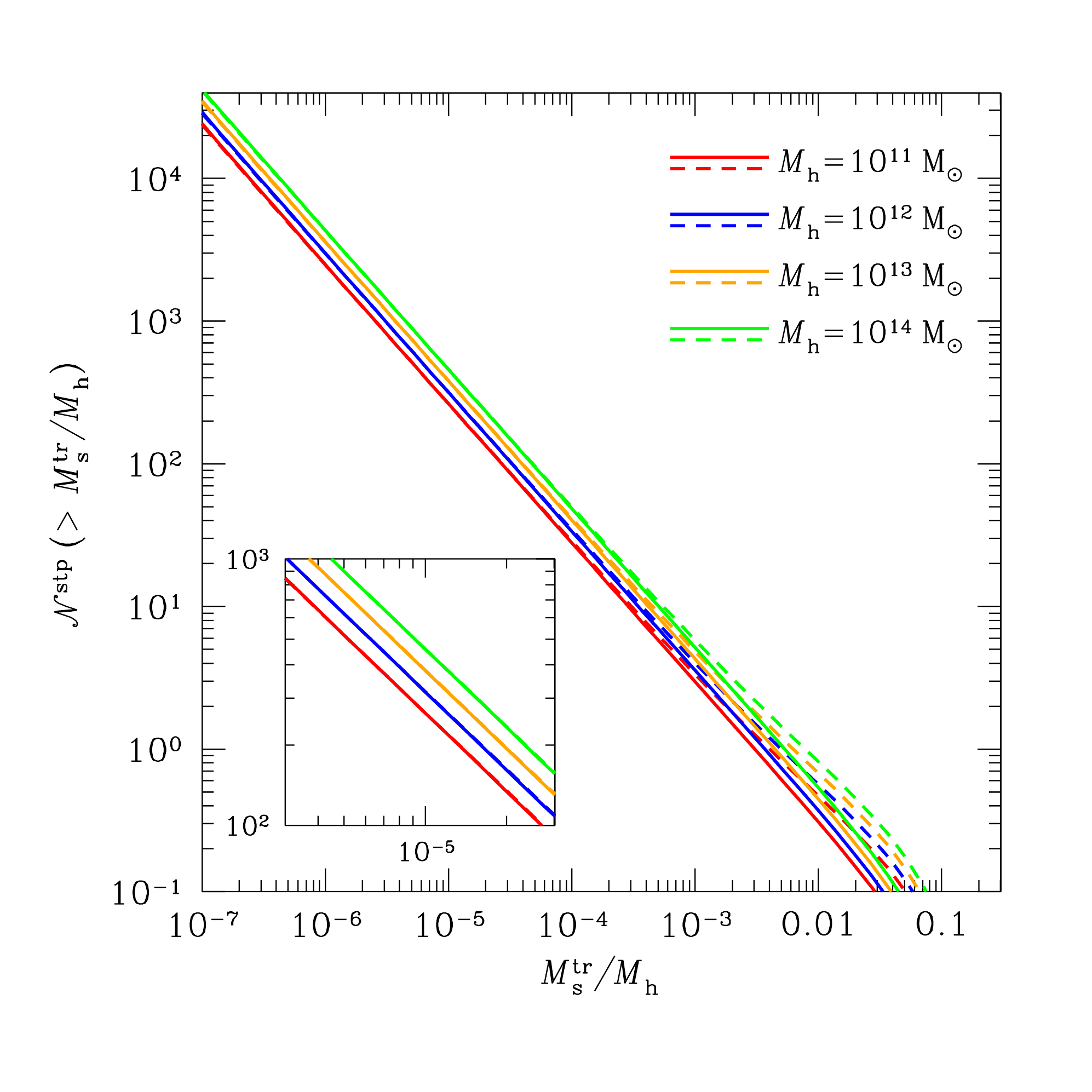}}
\caption{Cumulative MF of stripped subhaloes predicted for several halo masses (coloured lines). We plot the results obtained from our original stripping model with mass-dependent subhalo concentrations according to the Gao et al. $M$--$c$ relation (solid lines) and the corrected and renormalized version with suppression of stripping when the mass of subhaloes is larger than $M(r\per)$ (dashed lines).}
(A colour version of this Figure is available in the online journal.)
\label{f10}
\end{figure}

We thus see that accounting for the mass dependence of halo concentration is not enough to reproduce the detailed differential MF of stripped subhaloes. To do that we need, in addition, to include the effects of dynamical friction. Nevertheless, our model finds the dependence on halo mass of the MF of stripped subhaloes (the MF of accreted subhaloes is universal; see Paper I). This result, which is the direct consequence of the dependence of stripping on halo concentration, is very robust as it is little sensitive to the particular $M$--$c$ relation adopted. This success thus gives strong support to the central role of halo concentration in subalo stripping as considered in our model (see also \citealt{Chea17}). This would be the origin of the mass dependence of the cumulative MFs of stripped subhaloes (see Fig.~\ref{f10}) found in simulations (\citealt{Zeabis05,Gi08,Gea11,Iea20}).

\section{SUMMARY AND DISCUSSION}\label{dis}

Using the results of Paper I on accreted subhaloes and dDM, we have calculated the typical (mean) abundance and radial distribution of stripped subhaloes and dDM in current MW-mass haloes in a $\Lambda$CDM cosmology. To do this we have modelled the repetitive tidal stripping and heating suffered by subhaloes orbiting within purely accreting host haloes ignoring the effects of dynamical friction. 

Contrarily to the derivation followed in Paper I, which was achieved from first principles and with no single free parameter, that followed here has involved a model of subhalo tidal heating with two free parameters. We remark, however, that these parameters have not been tuned to get a good fit to the final properties of substructure found in simulations, but just to fit the effect of repetitive stripping and heating found in dedicated numerical experiments. As the general conclusions reached do not depend on the particular values of these parameters, we can thus still say that the derivation followed is parameter-free.

Our detailed treatment has payed special attention to the role of the dDM and subhaloes previously locked within accreted subhaloes and released in the intra-halo medium through the stripped material. The proportion of stripped subhaloes arising from released subsubhaloes per each accreted subhalo is only of 6 \% at the outer radius where it reaches the maximum value in current MW-mass haloes. Thus, this contributions to the total abundance of stripped subhaloes can be neglected in a first approximation. In respect to the proportion of released to accreted dDM, it starts increasing with increasing halo radius at $r\sim 2\times 10^{-4} R\h$ ($12\times 10^{-3} R\h$), reaches a maximum of about 3 (1.7) at $r\sim 0.1 R\h$ (0.1$R\h$) and then decreases until a value of 2 (1) at the virial radius of current MW-mass haloes in a real 100 GeV WIMP universe (and a SWV-like simulation). This represents a total dDM mass fraction of $\sim 0.95$ in both cases. In other words, the structure of current haloes is amply dominated, particularly in the central region, by dDM, which has important consequences for the boost factor of the theoretical DM annihilation signal. 

We have clarified the origin of HCFJ conditions shown to encode the properties of substructure. In Paper I the conditions 1 and 2 were demonstrated to arise directly from the properties of peaks in the Gaussian random field of density perturbations. In the present Paper we have shown that condition 3 stating that the truncated-to-original subhalo mass ratio profile does not depend on subhalo mass arises from the similar concentrations of accreted subhaloes, together with the fact that subhaloes suffer maximum tidal stripping at pericentre, where they are truncated at the radius encompassing a mean inner density of the order of that of the host halo there. However, according to our results, HCFJ condition 3 is only approximate because of the weak but non-negligible dependence of subhalo concentration on mass. Nonetheless, this dependence has little effect on the scaled subhalo number density profile, which is kept essentially independent of subhalo mass as found in simulations (provided dynamical friction is ignored; \citep{Hea18}). 

Given the dependence of subhalo concentration on mass, our detailed quantitative results depend on the exact $M$--$C$ relation assumed. When we use the empirical $M$--$c$ relation found in simulations by \citet{Gea08}, our predicted scaled number density profile of stripped subhaloes fully reproduces the empirical one foiund by HCFJ in the Aquarius simulation halo A, which is substantially shallower than the density profile of the host halo. This is so despite that the predicted mean truncated-to-original subhalo mass ratio is substantially lower than the corresponding empirical median profile derived by HCFJ. This result, which is contrary to the expectations for the lognormal distribution of truncated-to-original mass ratios, is likely due to the fact that the median profile obtained by HCFJ has been derived for subhaloes of all levels, while our predictions are for first-level ones only. Since the number of subhaloes of any mass at all levels is twice that of first-level subhaloes (Paper I), the only ones undergoing stripping, it is not unsurprising that the truncated-to-original profile for subhaloes of the former population is notably higher than for the latter one, while they both have the same {\it scaled} number density profile. When the unbiased $M$--$c$ relation predicted by CUSP is used, the predicted mean truncated-to-original subhalo mass ratio somewhat changes, but the general trend is similar. In particular, the corresponding scaled number density profile becomes substantially steeper, but it is also kept less steep than the mass density profile of the halo. This robust result is the consequence of the higher concentration of dDM towards the halo centre. On the other hand, the predicted subhalo MF reproduces the subhalo MF and its dependence on halo mass found in simulations, regardless of the particular $M$--$c$ relation used. 

All these results have been derived neglecting the effects of dynamical friction. This is an important limitation for stripped subhaloes with masses above $10^{-4}M\h$, whose radial distribution and MF are notably affected by that mechanism. This is the reason why we are currently working in the implementation of an accurate analytic treatment of dynamical friction in the stripping model.

On the other hand, the results presented hold for haloes having grown by pure accretion. As shown in \citep{SM19}, the violent relaxation that takes place in major mergers causes the system to lose the memory of its past assembly history, so the general properties inferred for purely accreting haloes also hold for ordinary haloes having suffered major mergers. However, that general rule does not hold for stripped subhaloes because the imprints of stripping are not erased by violent relaxation. Thus, even though the results found here for purely accreting haloes reproduce the properties of substructure found in the Level 1 Aquarius halo A having been accreting since $r\sim 0.08R\h$ and also seem to agree with the properties of substructure found in simulated haloes in general, we cannot discard that such properties slightly depend on the merger history of haloes. The more realistic case of haloes having suffered major mergers is addressed in Paper III, where we focus on very low-mass subhaloes whose properties should not be influenced by dynamical friction. 

\par\vspace{0.75cm}\noindent
{\bf DATA AVAILABILITY}

\vspace{11pt}\noindent 
The data underlying this article will be shared on reasonable request to the corresponding author.

\vspace{0.75cm} \par\noindent
{\bf ACKNOWLEDGEMENTS} 

\vspace{11pt}\noindent
One of us, I.B., has benefited of a MEXT scholarship by the Japanese MECSST. Funding for this work was provided by the Spanish MINECO under projects CEX2019-000918-M of ICCUB (Unidad de Excelencia `Mar\'ia de Maeztu') and PID2019-109361GB-100 (this latter co-funded with FEDER funds) and the Catalan DEC grant 2017SGR643.

{}

%% The appendix
%\clearpage

\appendix

\onecolumn

\section{Fraction of Released Subsubhaloes}\label{App1}

Dividing equation (\ref{corrbis}) by ${\cal N}\tr(r,\clM\tr)$ and defining $1+f\trb(r,\clM\tr)$ as ${\cal N}\fin(r,\clM\tr)/{\cal N}\tr(r,\clM\tr)$, we have
\beq 
1+f\trb(r,\clM\tr)=
1+\int_{\clM}^{M(r)} \,\der M\, {\cal N}\acc(r,
M)\left\lav \int_{R\tr(v,r)}^{R(r,M)}\der r'\,
\frac{{\cal N}\fin_{\rm [M,t(r)]}(r',\clM\tr)}{{\cal N}\tr_{\rm [M,t(r)]}(r',\clM)}\,\frac{{\cal N}\tr_{\rm [M,t(r)]}(r',\clM)}{{\cal N}\tr(r,\clM\tr)}
\right\rav\,.
\label{corr00} 
\eeq
The quantities ${\cal N}\fin_{\rm [M,t(r)]}(r',\clM\tr)$ and ${\cal N}\tr_{\rm [M,t(r)]}(r',\clM\tr)$ inside the integral on the right can be written in terms of the quantities $f\trb(r')$ and $\rho(r')$ with subindex $[M,t(r)]$. All these quantities seen as functions of the scaled radius $\xi=r'/R(r,M)$ are universal, so we can change the subindex $[M,t(r)]$ by subindex $[M(r),t(r)]$ and, given the inside-out growth of the halo, by subindex $[M\h,t\h]$ or simply drop the subindex. (Of course, in this case, when rewriting the latter quantity as a function of the non-scaled radius, $\xi$ must be multiplied by $R\h$.) After all those changes we have 
\beq 
1+f\trb(r,\clM\tr)=1+\,\int_{\clM}^{M(r)}\der M\, {\cal N}\acc(r,
M) \frac{R^3(r,M)}{r^2}\left\lav \int_{Q_{\nu(v,r)}(v,r)}^1\der \xi\,[1+f\trb(\xi,{\clM\tr})]\,\xi^2\,\frac{\mu(\xi)}{\mu(r)}\frac{\rho(\xi)}{\rho(r)}\right\rav\,.
\label{corr01} 
\eeq
Taking into account the relation (\ref{ratio}) and the equality $R^3(r,M)=3M/[4\pi \bar\rho(r)]$ (eq.~[3]), equation (\ref{corr01}) can be written as
\beq 
f\trb(r,\clM\tr)\mu(r)=\frac{M(r)-\clM}{M\h\bar\rho(r)}\left\lav 3\int_{Q_{\nu(v,r)}(v,r)}^1\der \xi\,\xi^2\,[1+f\trb(\xi,\clM\tr)]\,\mu(\xi)\,\rho(\xi)\right\rav\,.
\label{corr000} 
\eeq
For any given subhalo mass $\clM\tr$, equation (\ref{corr000}) holds for $r$ satisfying the condition $\clM\tr\le M(r)/3$. Near equality, where ${\cal N}\acc(r,\clM\tr)$ vanishes, $f\trb(r,\clM\tr)$ falls rapidly to zero and, over all the remaining radii, $\clM$ in the denominator on the right of equation (\ref{corr000}) can be neglected.Thus, $f\trb(r,\clM\tr)$ coincides for all subhalo masses $\clM\tr$ and we can drop its argument $\clM\tr$. Lastly, by performing through partial integration the average over $v$ on the right side of equation (\ref{corr000}), we arrive at equation (\ref{final3}).

\section{Stripping of Released Subsubhaloes}\label{App2}

As shown in Section \ref{stripped}, the average over $v$ of the Jacobian $\partial \clM\tr(v,r,\clM)/\partial \clM$ is equal to the average over $v$ of the $M\tr(v,r,\clM)/\clM$ mass ratio, which only depends on $r$. Taking into account this relation for subsubhaloes with truncated mass $\clM\tr$ released from first-level subhaloes into the intra-halo medium, we arrive, after integrating over the velocities $v'$ in the host subhalo and $v$ in the host halo at the following condition
for a new tidal stripping to take place,
\beq
M\trh= \left\lav\left\lav\frac{\partial M\trh}{\partial M\trs}\right\rav\right\rav\!(r,r')\, M\trs=\left\lav\frac{\partial M\trh}{\partial M}\right\rav\!(r)\left\lav\frac{\partial M}{\partial M\trs}\right\rav\!(r')\,M\trs\,,
\label{mtrbis2}
\eeq
where the extra index h or s going together with superindex s means that truncation takes place within the halo or the subhalo, respectively, $r'$ is the apocentric radius of the subsubhalo within the host subhalo, and $r$ is the apocentric radius of that subhalo (and of the released subsubhalo)
within the host halo. The condition for a new tidal stripping is thus
\beq 
\left\lav\frac{\partial \clM\tr}{\partial 
  \clM}\right\rav\!(r) < \left\lav\frac{\partial \clM\tr}{\partial
  \clM}\right\rav\!(r')\,.
\label{con}
\eeq
The two partial derivatives are independent of the subsubhalo mass $\clM$. Since they refer to different hosts (to the halo that on the left and to the stripped subhalo that on the right), condition (\ref{con}) seems hard to assess. In the hypothetical case that the host subhalo accreted at $t(r)$ were identical to its host at that moment, the two partial derivatives would correspond to the same host at different radii, with $r'$ smaller than $r$. Since stripping is more intense near the halo centre ($c(r)=c(M\h,t\h)\,R\h/r$), the partial derivative on the right (and the associated $v$-averaged mass ratio) would be smaller than that on the left. Consequently, condition (\ref{con}) would not be satisfied and there would be no new tidal stripping within the host halo (although there would still be repetitive stripping from that initial stripped configuration). Actually, accreted subhaloes are always less massive than the accreting host at the time of their accretion, so subhaloes are necessarily (slightly) more concentrated than the host and the stripping the subsubhalo suffers is stronger than the one it would suffer were the subhalo identical to the host at accretion. Therefore, there is no additional initial tidal stripping in the realistic case (just the usual repetitive stripping).

This conclusion holds, however, after averaging over $v$ and $v'$. In the case that $v'$ is very large, causing a very small stripping of the subsubhalo inside the subhalo host, and $v$ is very small, causing a very marked stripping of the released subsubhalo inside the host halo, subsubhaloes will suffer a new stripping after being released in the intra-halo medium. But such a configuration should also likely cause the disruption of the released subhalo so that surviving subhaloes having undergone a new stripping after being released from
the stripping of other subhaloes are expected to be very rare.

\section{Tangential velocity distribution function}\label{veldis}

CUSP allows deriving the total (3D), radial and tangential velocity dispersion profiles of haloes (SSMG), but not their respective velocity distributions. Thus, to perform some explicit calculations, the averages over the tangential velocity of (accreted or stripped) subhaloes with apocentre at $r$ have been performed using the (mass-independent; \citealt{Jea15}) tangential velocity distribution function of the \citet{T88} form,
\beq
f\tang(v,r)\propto v\left[1+\frac{8v^2}{3\sigma^2(r)}\right]^{-5/2},
\label{Tsallis}
\eeq
where $\sigma(r)$ is the 3D velocity dispersion profile, found for all particles in simulated haloes \citep{Hea06}.

Of course, this empirical distribution function must be adapted to our needs because it refers to all particles at $r$, while what we need is the distribution function for particles with apocentre at that radius. As orbiting particles spend most of the time near apocentre, the particles we are interested in dominate by far the total population at $r$. There are just a few more particles caught when they are crossing $r$ from larger apocentric radii (i.e. they belong to accreted shells that were not yet virialised at the time $t(r)$). Although they are not numerous, these particles give rise to the otherwise null radial velocity dispersion at $r$. They also fill the tangential velocity distribution function (\ref{Tsallis}) beyond $v\maxi$, a region inaccessible to particles with apocentre at $r$, and likely also have a substantial contribution to that distribution function near $v\maxi$ where the subhalo population with apocentre at $r$ is small \citep{T97,Zeabis05,W11}. However, at lower tangential velocities, particles with apocentre at $r$ should clearly predominate. Therefore, the velocity distribution of subhaloes and DM particles with apocentre at $r$ should be well approximated by the Tsallis distribution function (\ref{Tsallis}) convolved with a Gaussian with central value equal to unity and 3-$\sigma$ equal to $v\maxi$. This is the approximate form we adopt. Remember that, when the possibility of disruption is considered, the velocity distribution used to average over all {\it surviving} stripped subhaloes must also be taken null (in this case with a sharp cutoff) for $v< v\des$.

Regarding the 3D velocity dispersion profile appearing in expression (\ref{Tsallis}), we could use that predicted by CUSP (SSMG). However, that velocity dispersion was derived for realistic triaxial haloes, while we are assuming here spherical symmetry. Thus we simply adopt the solution of the isotropic Jeans equation for spherically symmetric haloes endowed with the NFW profile with null boundary condition at infinity \citep{CL96}.


\begin{thebibliography}{}
%
\bibitem[Angulo et al.(2009)]{Aea09} Angulo R.~E., Lacey C.~G., Baugh C.~M., Frenk C.~S., 2009, \mnras, 399, 983
%
\bibitem[Angulo \& White(2010)]{AW10} Angulo R.~E. \& White S.~D.~M.\ 2010, \mnras, 401, 1796 
%
\bibitem[Bardeen et al.(1986)]{BBKS} Bardeen J.~M., Bond J.~R., Kaiser N., Szalay A.~S., 1986, \apj, 304, 15
%
\bibitem[Benson et al.(2013)]{Bea13} Benson A.~J., Farahi A., Cole S., et al., 2013, \mnras, 428, 1774
%
\bibitem[Binney \& Tremaine(2008)]{BT08} Binney J. \& Tremaine J., 2008, Galactic dynamics: Second Edition. Princeton University Press
%
\bibitem[\protect\citeauthoryear{Bose et al.}{2016}] {Bea16} Bose S., Hellwing W.~A., Frenk C.~S., Jenkins A., Lovell M.~R., Helly J.~C., Li B., 2016, MNRAS, 455, 318%. doi:10.1093/mnras/stv2294
%
\bibitem[\protect\citeauthoryear{Bose et al.}{2020}]{Bea20} Bose S., Deason A.~J., Belokurov V., Frenk C.~S., 2020, MNRAS, 495, 743%. doi:10.1093/mnras/staa1199
%
\bibitem[\protect\citeauthoryear{Boylan-Kolchin et al.}{2010}]{BK10} Boylan-Kolchin M., Springel V., White S.~D.~M., Jenkins A., 2010, MNRAS, 406, 896%. doi:10.1111/j.1745-3933.2011.01074.x
%
\bibitem[\protect\citeauthoryear{Bryan \& Norman}{1998}]{BN98} Bryan G.~L., Norman M.~L., 1998, ApJ, 495, 80%. doi:10.1086/305262
%
\bibitem[Cautun et al.(2014)]{Cea14} Cautun M., Frenk C.~S., van de Weygaert R., Hellwing W.~A., Jones B.~J.~T., 2014a, \mnras, 445, 2049
%
\bibitem[\protect\citeauthoryear{Chua et al.}{2017}]{Chea17} Chua K.~T.~E., Pillepich A., Rodriguez-Gomez V., Vogelsberger M., Bird S., Hernquist L., 2017, MNRAS, 472, 4343%. doi:10.1093/mnras/stx2238
%
\bibitem[Cole \& Lacey(1996)]{CL96} Cole S. \& Lacey C.,1996, \mnras, 281, 716 
%
\bibitem[Diemand, Moore, \& Stadel(2004)]{Dea04} Diemand J., Moore B., Stadel J., 2004, MNRAS, 353, 624. %doi:10.1111/j.1365-2966.2004.08094.x
%
\bibitem[Diemand et al.(2007)]{Dea07} Diemand J., Kuhlen M., Madau P., 2007, \apj, 657, 267
%
\bibitem[Einasto(1965)]{E65} Einasto J., 1965, Trudy Inst. Astrofiz. Alma-Ata, 5, 87
%
\bibitem[Elahi et al.(2009)]{Eea09} Elahi P.~J., Widrow L.~M., Thacker R.~J., 2009, Ph.~Rev.~D, 80, 123513
%
\bibitem[\protect\citeauthoryear{Errani \& Pe{\~n}arrubia}{2020}]{EP20} Errani R., Pe{\~n}arrubia J., 2020, MNRAS, 491, 4591%. doi:10.1093/mnras/stz3349
%
\bibitem[\protect\citeauthoryear{Fielder et al.}{2020}]{Fiea20} Fielder C.~E., Mao Y.-Y., Zentner A.~R., Newman J.~A., Wu H.-Y., Wechsler R.~H., 2020, MNRAS, 499, 2426%. doi:10.1093/mnras/staa2851
%
\bibitem[\protect\citeauthoryear{Font et al.}{2020}]{Fea20} Font A.~S., McCarthy I.~G., Poole-Mckenzie R., Stafford S.~G., Brown S.~T., Schaye J., Crain R.~A., et al., 2020, MNRAS, 498, 1765%. doi:10.1093/mnras/staa2463
%
\bibitem[\protect\citeauthoryear{Font, McCarthy, \& Belokurov}{2020}]{Fea20b} Font A.~S., McCarthy I.~G., Belokurov V., 2020, arXiv, arXiv:2011.12974
%
\bibitem[Fujita et al.(2002)]{Fea02} Fujita Y., Sarazin C.~L., Nagashima M., Yano T., 2002, \apj, 577, 11
%
\bibitem[Gao et al.(2004)]{Gea04} Gao L., White S.~D.~M., Jenkins A., Stoehr F., Springel V., 2004, \mnras, 355, 819%. doi:10.1111/j.1365-2966.2004.08360.x
%
\bibitem[\protect\citeauthoryear{Gao et al.}{2008}]{Gea08} Gao L., Navarro J.~F., Cole S., Frenk C.~S., White S.~D.~M., Springel V., Jenkins A., et al., 2008, MNRAS, 387, 536%. doi:10.1111/j.1365-2966.2008.13277.x
%
\bibitem[Gao et al.(2011)]{Gea11} Gao L., Frenk C.~S., Boylan-Kolchin M., Jenkins A., Springel V., White S.~D.~M., 2011, \mnras, 410, 2309
%
\bibitem[Gao et al.(2012)]{Gea12} Gao L., Frenk C.~S., Jenkins A., Springel V., White S.~D.~M., 2012, \mnras, 419, 1721
%
\bibitem[Ghigna et al.(1998)]{Gea98} Ghigna S., Moore B., Governato F., Lake G., Quinn T., Stadel J., 1998, \mnras, 300, 146
%
\bibitem[Giocoli et al.(2008)]{Gi08} Giocoli C., Tormen 
G., van den Bosch F.~C., 2008, \mnras, 386, 2135 
%
\bibitem[Giocoli et al.(2010)]{Gi10} Giocoli C., Tormen G., Sheth R.~K., van den Bosch F.~C., 2010, \mnras, 404, 502
%
\bibitem[Gnedin \& Ostriker(1999)]{GO99} Gnedin O.~Y. \& Ostriker J.~P., 1999, \apj, 513, 626
%
\bibitem[Gonz\'alez-Casado at al.(1994)]{Gea94} Gonz\'alez-Casado G., Mamon G.~A., Salvador-Sol\'e E., 1994, \apjl, 433, L61
%
\bibitem[Green \& van den Bosch(2019)]{GB19} Green S.~B., van den Bosch F.~C., 2019, MNRAS, 490, 2091 %. doi:10.1093/mnras/stz2767
%
\bibitem[Griffen et al.(2016)]{Gfea16} Griffen B.~F., Ji A.~P., Dooley G.~A., G{\'o}mez F.~A., Vogelsberger M., O'Shea B.~W., Frebel A., 2016, ApJ, 818, 10%. doi:10.3847/0004-637X/818/1/10
%
\bibitem[\protect\citeauthoryear{Han et al.}{2012}]{Hea12} Han J., Jing Y.~P., Wang H., Wang W., 2012, MNRAS, 427, 2437%. doi:10.1111/j.1365-2966.2012.22111.x
%
\bibitem[Han et al.(2016)]{Han15} Han J., Cole S., Frenk C.~S., Jing Y., 2016, \mnras, 457, 1208 (HCFJ)
%
\bibitem[\protect\citeauthoryear{Han et al.}{2018}]{Hea18} Han J., Cole S., Frenk C.~S., Benitez-Llambay A., Helly J., 2018, MNRAS, 474, 604%. doi:10.1093/mnras/stx2792
%
\bibitem[Hansen et al.(2006)]{Hea06} Hansen S.~H., Moore B., Zemp M., \& Stadel, J., 2006, JCAP, 1, 014 
%
\bibitem[Hayashi et al.(2003)]{Hay03} Hayashi E., Navarro 
 J.~F., Taylor J.~E., Stadel J., Quinn T., 2003, \apj, 584, 541 
%
\bibitem[Hellwing et al.(2016)]{Hell16} Hellwing W.~A., Frenk C.~S., Cautun M., Bose S., Helly J., Jenkins A., Sawala T., et al., 2016, MNRAS, 457, 3492%. doi:10.1093/mnras/stw214
%
\bibitem[Henry(2000)]{H00} Henry, J.~P., 2000, \apj, 534, 565 
%
\bibitem[\protect\citeauthoryear{Ishiyama et al.}{2020}]{Iea20} Ishiyama T., Prada F., Klypin A.~A., Sinha M., Metcalf R.~B., Jullo E., Altieri B., et al., 2020, arXiv, arXiv:2007.14720
%
\bibitem[Jiang \& van den Bosch(2016)]{Ji16} Jiang F. \& van den Bosch F.~C.\ 2016, \mnras, 458, 2848
%
\bibitem[Jiang et al.(2015)]{Jea15} Jiang L., Cole S., Sawala T., Frenk C.~S., 2015, MNRAS, 448, 1674%. doi:10.1093/mnras/stv053
%
\bibitem[\protect\citeauthoryear{Jiang et al.}{2021}]{Jea21} Jiang F., Dekel A., Freundlich J., van den Bosch F.~C., Green S.~B., Hopkins P.~F., Benson A., et al., 2021, MNRAS, 502, 621%. doi:10.1093/mnras/staa4034
%
\bibitem[Juan et al.(2014a)]{Jea14a} Juan E., Salvador-Sol\'e E., Dom\`enec G., Manrique A., 2014, \mnras, 439, 719
%
\bibitem[Juan et al.(2014b)]{Jea14b} Juan E., Salvador-Sol{\'e} E., Dom{\`e}nech G., Manrique A., 2014, \mnras, 439, 3156
%
\bibitem[Kampakoglou \& Benson(2007)] {KB07} Kampakoglou M. \& Benson A.~J., 2007, \mnras, 374, 775 
%
\bibitem[King(1962)]{K62} King I., 1962, \aj, 67, 471 
%
\bibitem[Klypin et al.(1999)]{Kea99} Klypin A., Gott\"ober S., Kravtsov A.~V., 1999, \apj, 516, 530
%
\bibitem[Klypin et al.(2011)]{Kea11} Klypin A.~A., Trujillo-Gomez S., Primack J., 2011, \apj, 740, 102
%
\bibitem[Komatsu et al.(2011)]{Km11} Komatsu E., Smith K. M., Dunkley J., Bennet C. L., Gold B., Hinshaw G., Jarosik N., et al. others, 2011, ApJS, 192, 18
%
\bibitem[Lee(2004)]{L04} Lee J., 2004, \apj, 604, L73
%
\bibitem[Lovell et al.(2014)]{Lea14} Lovell M.~R., Frenk  C.~S., Eke V.~R., et al., 2014, \mnras, 439, 300 
%
\bibitem[Ludlow et al.(2009)]{Lea09} Ludlow A.~D., Navarro J.~F., Springel V., Jenkins A., Frenk C.~S., Helmi A., 2009, ApJ, 692, 931%. doi:10.1088/0004-637X/692/1/931
%
\bibitem[Manrique \& Salvador-Sol\'e(1995)]{Mea95} Manrique A. \& Salvador-Sol\'e E., 1995, \apj, 453, 6
%
\bibitem[Manrique \& Salvador-Sol\'e(1996)]{Mea96} Manrique A. \& Salvador-Sol\'e E., 1996, \apj, 467, 504
%
\bibitem[Manrique et al.(1998)]{Mea98} Manrique A., Raig A., Solanes J. M., Gonz\'alez-Casado G., Stein, P., Salvador-Sol\'e E., 1998, \apj, 499, 548
%
\bibitem[Mo et al.(2010)]{Mea10} Mo H., van den Bosch F.~C., White S., 2010, Galaxy Formation and Evolution. Cambridge University Press
%
\bibitem[Nagai \& Kravtsov(2005)]{NK05} Nagai D., Kravtsov A.~V., 2005, ApJ, 618, 557. doi:10.1086/426016
%
\bibitem[Navarro et al.(1997)]{NFW97} Navarro J.~F., Frenk C.~S., White S.~D.~M., 1997, \apj, 490, 493
%
\bibitem[Oguri \& Lee(2004)]{OL04} Oguri M. \& Lee J. 2004, \mnras, 355, 120
%
\bibitem[Onions et al.(2012)] {Oea12} Onions J., Knebe A., Pearce F.~R., Muldrew S.~I., Lux H., Knollmann S.~R., Ascasibar Y., et al., 2012, MNRAS, 423, 1200%. doi:10.1111/j.1365-2966.2012.20947.
%
\bibitem[Pe{\~n}arrubia et al.(2008)]{Pea08} Pe{\~n}arrubia, J., Navarro J.~F., McConnachie A.~W., 2008, \apj, 673, 226-240 
%
\bibitem[Pe{\~n}arrubia \& Benson(2005)]{Pe05} Pe{\~n}arrubia J. \& Benson A.~J., 2005, \mnras, 364, 977
%
\bibitem[Pullen et al.(2014)]{PB14} Pullen A.~R., Benson A.~J., Moustakas L.~A., 2014, \apj, 792, 24 
%
\bibitem[Read et al.(2006)]{Rea06} Read J.~L., Wilkinson M.~I., Evans N.W., Gilmore G., Kleyna J.~T., 2006, \mnras, 366,429 
%
\bibitem[\protect\citeauthoryear{Richings et al.}{2020}]{Rich20} Richings J., Frenk C., Jenkins A., Robertson A., Fattahi A., Grand R.~J.~J., Navarro J., et al., 2020, MNRAS, 492, 5780%. doi:10.1093/mnras/stz3448
%
\bibitem[Rodr{\'\i}guez-Puebla et al.(2016)]{Rea16} Rodr{\'\i}guez-Puebla, A., Behroozi, P., Primack, J., et al.\ 2016, \mnras, 462, 893%. doi:10.1093/mnras/stw1705
%
\bibitem[Salvador-Sol{\'e} et al.(2012a)]{Sea12a} Salvador-Sol{\'e} E., Vi{\~n}as J., Manrique A., Serra S., 2012a, \mnras, 423, 2190
%
\bibitem[Salvador-Sol{\'e} et al.(2012b)]{Sea12b} Salvador-Sol{\'e} E., Serra S., Manrique A., Gonz{\'a}lez-Casado G., 2012b, \mnras, 424, 3129 (SSMG)
%
\bibitem[\protect\citeauthoryear{Salvador-Sol{\'e} \& Manrique}{2021}]{SM19} Salvador-Sol{\'e} E., Manrique A., 2021, \apj, 914,141
%
\bibitem[Salvador-Sol{\'e} et al.(2021a)]{I} Salvador-Sol{\'e} E., Manrique A., Botella I., 2021a, \mnras in press, arXiv:2109.06484 (Paper I)
%
\bibitem[Salvador-Sol{\'e} et al.(2021b)]{III} Salvador-Sol{\'e} E., Manrique A., Canales D., Botella I., 2021b, submitted to \mnras (Paper III)
%
\bibitem[Salvador-Sol{\'e} et al.(2021c)]{Jea19} Salvador-Sol{\'e} E., Canales D., Manrique A., Juan E. \& Botella, I.\ 2021c in preparation
%
\bibitem[Sheth (2003)]{S03} Sheth R.~K., 2003, \mnras, 345, 1200
%
\bibitem[Spitzer(1958)]{Sp58} Spitzer L., Jr., 1958, \apj, 127, 17 
%
\bibitem[Spitzer(1978)]{Sp87} Spitzer L., Jr., 1987, Dynamical evolution of globular clusters. Princeton University Press 
%
\bibitem[Springel et al.(2008)]{Sea08a} Springel V., Wang J., Vogelsberger M., et al., 2008a, \mnras, 391, 1685 (SWV)
%
\bibitem[\protect\citeauthoryear{Sugiyama}{1995}]{S95} Sugiyama, N.\ 1995, \apjs, 100, 281 
%
\bibitem[Taylor \& Babul(2001)]{TB01} Taylor J.~E. \& Babul A., 2001, \apj, 559, 716
%
\bibitem[Taylor \& Babul(2004)]{TB04} Taylor J.~E. \& Babul A., 2004, \mnras, 348, 811
%
\bibitem[Tollet et al.(2017)]{Tea17} Tollet {\'E}., Cattaneo A., Mamon G.~A., Moutard T., van den Bosch F.~C., 2017, MNRAS, 471, 4170%. doi:10.1093/mnras/stx1840
%
\bibitem[Tormen(1997)]{T97} Tormen G., 1997, \mnras, 290, 411
%
\bibitem[Tormen et al.(1998)]{Tea98} Tormen G., Diaferio A., Syer D., 1998, \mnras, 299, 728
%
\bibitem[Tsallis(1988)]{T88} Tsallis C., 1988, J. Stat. Phys., 52, 479
%
\bibitem[Vi{\~n}as et al.(2012)]{Vea12} Vi{\~n}as J., 
Salvador-Sol{\'e} E., Manrique A., 2012, \mnras, 424, L6 
%
\bibitem[van den Bosch et al.(2005)]{vdB05} van den Bosch, F.~C., Tormen G., Giocoli C., 2005, \mnras, 359, 1029 
%
\bibitem[van den Bosch \& Jiang(2016)]{vdB16} van den Bosch F.~C. \& Jiang F., 2016, MNRAS, 458, 2870%. doi:10.1093/mnras/stw440
%
\bibitem[van den Bosch et al.(2018)]{vdB18} van den Bosch F.~C., Ogiya G., Hahn O., Burkert A., 2018, \mnras, 474, 3043%. doi:10.1093/mnras/stx2956
%
\bibitem[Wang et al.(2011)]{Wa11} Wang J., Navarro J.~F., Frenk C.~S., et al., 2011, \mnras, 413, 1373
%
\bibitem[Weinberg(1994)]{W94} Weinberg M.~D., 1994, \aj, 108, 1398
%
\bibitem[Wetzel(2011)]{W11} Wetzel A.~R., 2011, \mnras, 412, 49
%
\bibitem[Zentner \& Bullock(2003)]{ZB03} Zentner A.~R., Bullock, J.~S., 2003, \apj, 598, 49
%
\bibitem[Zentner et al.(2005)]{Zeabis05} Zentner A.~R., Berlin A.~A., Bullock J.~S., Kravtsov A.~V., Wechsler R.~H., 2005, \apj, 624, 505
%
\end{thebibliography}
\end{document}